\documentclass[a4paper,fleqn,usenatbib]{mnras}

\usepackage[T1]{fontenc}
\usepackage{ae,aecompl}

\usepackage{graphicx}
\usepackage{amsmath}
\usepackage{stmaryrd}

\newcommand{\simnot}{\mathord{\sim}}
\mathchardef\mhyphen="2D

\title[The hard and intermediate states of XRBs]{Probing the hard and intermediate states of X-ray binaries using short time-scale variability}

\author[C.~J.~Skipper and I.~M.~M\textsuperscript{c}Hardy]{
Chris~J.~Skipper$^{1,2}$
and Ian~M.~M\textsuperscript{c}Hardy$^1$
\\
$^{1}$School of Physics and Astronomy, University of Southampton, Southampton, SO17 1BJ, UK\\
$^{2}$Jodrell Bank Centre for Astrophysics, Alan Turin Building, The University of Manchester, Manchester, M13 9PL, UK
}

\date{Accepted XXX. Received YYY; in original form ZZZ}

\pubyear{2016}

\begin{document}
\label{firstpage}
\pagerange{\pageref{firstpage}--\pageref{lastpage}}

\maketitle

\begin{abstract}
Below an accretion rate of approximately a few per cent of the Eddington accretion rate, X-ray binary systems are not usually found in the soft spectral state. However, at accretion rates a factor of a few lower still, in the hard state, there is another spectral transition which is well observed but not well understood. Below $\simnot0.5$-1 per cent of the Eddington accretion rate ($\dot{m}_{\rm crit}$), the spectral index hardens with increasing accretion rate, but above $\dot{m}_{\rm crit}$, although still in the hard state, the spectral index softens with increasing accretion rate. Here we use a combination of X-ray spectral fitting and a study of short time-scale spectral variability to examine the behaviour of three well-known X-ray binaries: Cygnus X-1, GX 339-4 and XTE J1118+480.

In Cygnus X-1 we find separate hard and soft continuum components, and show using root-mean-square (rms) spectra that the soft component dominates the variability. The spectral transition at $\dot{m}_{\rm crit}$ is clearly present in the hard-state hardness-intensity diagrams of Cygnus X-1. Above $\dot{m}_{\rm crit}$, GX 339-4 shows similar softer-when-brighter behaviour at both long and short time-scales. Similarly, XTE J1118+480, which remains well below $\dot{m}_{\rm crit}$, is harder-when-brighter behaviour on all time-scales.

We interpret these results in terms of two continuum components: a hard power-law which dominates the spectra when the accretion rate is low, probably arising from Comptonisation of cyclo-synchrotron photons from the corona, and a soft power-law which dominates at higher accretion rates, arising from Comptonisation of seed photons from the accretion disc.
\end{abstract}

\begin{keywords}
X-rays: binaries - accretion, accretion discs - X-rays: individual: Cygnus X-1 - X-rays: individual: GX 339-4 - X-rays: individual: XTE J1118+480.
\end{keywords}

%%%%%%%%%%%%%%%%%%%%%%%%%%%%%%%%%%%%%%%%%%%%%%%%%%

%%%%%%%%%%%%%%%%% BODY OF PAPER %%%%%%%%%%%%%%%%%%

\section{Introduction}

Black hole X-ray binaries (XRBs) are usually observed in one of a number of characteristic states, each of which is associated with specific X-ray spectral and timing properties \cite[e.g.][]{Homan2005,Remillard2006,Belloni2010}. The hard (or low) state is believed to be the result of thermal Comptonisation of lower energy seed photons in a cloud of hot electrons, and the X-ray spectra are usually consistent with a model comprising a power-law (which takes the form $N_{E} \propto E^{-\Gamma}$, where $N_{\rm E}$ is the photon number density at energy $E$ and $\Gamma$ is the photon index), accompanied by a weak iron line at $\rm\simnot6.4~keV$ and a weak reflection component \cite[e.g.][]{Zdziarski1998,Ueda1994}.

In the soft (or high) state the X-ray spectra become dominated below $\rm\simnot10~keV$ by the thermal emission from a geometrically thin, optically thick accretion disc \citep{Shakura1973}, and the iron line and reflection components increase in strength. A power-law component, thought to be the result of Comptonisation of disc seed photons in a hybrid (thermal plus non-thermal) corona of high-energy electrons \citep{Sunyaev1980}, forms a tail out to $\rm\simnot1~MeV$ \citep{Grove1998}.

During the outburst cycle of a typical XRB the source moves from the hard state to the soft state and then back again, but in order to do so it must pass through the intermediate state. Although the intermediate state phase lasts for a relatively short period of time, the changes in the spectral and timing properties have led to this state being divided into the hard-intermediate state and the soft-intermediate state \cite[e.g.][]{Belloni2005-2,Belloni2006}.

The photon index of the power-law spectral component is known to be correlated with the source accretion rate ($\dot{m}$), or, alternatively, the X-ray luminosity, which is often used as a proxy for $\dot{m}$. However, the nature of this correlation is fundamentally different depending upon whether $\dot{m}$ is above or below a critical accretion rate ($\dot{m}_{\rm crit}$, typically 0.5-1 per cent of the Eddington limit). When $\dot{m} > \dot{m}_{\rm crit}$ then $\Gamma$ and $\dot{m}$ are positively correlated (the source is softer when brighter), and when $\dot{m} < \dot{m}_{\rm crit}$ then the same two properties are anti-correlated \citep{Sobolewska2011}. The $\dot{m}/\Gamma$ correlation has been established for XRBs on a variety of time-scales \citep{Wu2008,Zdziarski2002,Wu2010}, and also for active galactic nuclei (AGN) in large surveys \citep{Constantin2009, Gu2009, Shemmer2006} and individual sources \citep{Lamer2003,Sobolewska2009,Emmanoulopoulos2012}.

The transition from the hard state to the soft state typically occurs at an accretion rate of a few per cent of the Eddington limit \citep{Maccarone2003,Gierlinski2006}, and during this transition $\Gamma$ will increase significantly from $\simnot$1.4-1.8 to $\simnot2.5$ and above. However, the switch from harder-when-brighter to softer-when-brighter behaviour at $\dot{m}_{\rm crit}$ is \textit{not} a state transition, and after $\dot{m}$ increases beyond the $\dot{m}_{\rm crit}$ boundary (which is typically a factor of a few lower in accretion rate than the hard-to-soft state transition) the source will initially remain in the hard state.

Physically, the harder-when-brighter behaviour at $\dot{m} < \dot{m}_{\rm crit}$ is usually attributed to the Comptonisation of cyclo-synchrotron seed photons in a geometrically thick, optically thin and radiatively inefficient accretion flow (RIAF; e.g. \citealt{Narayan1994,Abramowicz1995}), which contrast with the mostly thermal disc seed photons that are thought to supply the Comptonised power-law in the soft state. The soft-state power-law is expected to be softer when brighter due to Compton cooling of the scattering corona \citep[e.g.][]{Guilbert1982}, but to observe this effect in XRBs requires spectral fitting in order to separate the power-law component from the less variable \citep{Churazov2001} thermal disc emission which typically dominates the spectrum below $\simnot10$ keV.

However, in the brightest limits of the hard state, in which $\dot{m} > \dot{m}_{\rm crit}$, the clear softer-when-brighter behaviour suggests the presence of a cool accretion disc which contributes little to the energy spectra above 2 keV but is still able to provide sufficient thermal seed photons in order to cool the Comptonising corona. It has been suggested \citep[e.g.][]{Esin1997,Done1999,Tomsick2009} that in this state the inner edge of the accretion disc is truncated at some distance from the black hole, so that the geometry in the bright hard state can be described as an outer truncated disc and an inner RIAF. As the accretion rate of the source increases there will be a corresponding decrease in the truncation radius, thereby shifting the balance of seed photon supply from the RIAF to the disc. Conversely, other studies \cite[e.g.][]{Miller2006,Liu2007,Reis2010} have suggested that, due to the detection of broad iron lines, as well as evidence from spectral fitting of the disc component, the cool accretion disc persists to the innermost stable circular orbit even in the hard state.

If the truncated disc model is correct, then it may be reasonable to expect the X-ray continuum to be composed of two separate power-law component: one produced in the inner RIAF that is harder-when-brighter and another produced in the outer corona that is softer-when-brighter. \cite{Yamada2013} showed that their hard-state \textit{Suzaku} spectra of Cygnus X-1 can be described by such a two-component model. Numerous other studies have also found that the hard-state spectra of Cygnus X-1 are well described by a model comprised of two power-law components \cite[e.g.][]{Gierlinski1999,Bock2011,Wilms2006,Grinberg2013}, or alternatively that the spectra 'pivot' around $\simnot$10-20 keV \citep{Cui2002,Malzac2006}.

Luminosity and spectral changes can occur on time-scales much shorter than the integrated time of a typical 1 - 3 kilosecond (ks) observation by the Rossi X-ray Timing Explorer (\textit{RXTE}), and we therefore combine spectral fitting of X-ray data from \textit{RXTE} with an examination of short time-scale (100 milliseconds [ms] to a few kiloseconds) spectral variability to investigate whether or not the hypothesised hard and soft power-law components could be responsible for producing the change in spectral behaviour at $\dot{m}_{\rm crit}$. We also determine how these two components evolve as the source brightens from quiescence right through the hard and hard-intermediate states.

\section{Observations and data reduction}

\subsection{The sample}

In order to examine the behaviour of XRBs over a wide range of accretion rates, including both above and below the $\dot{m}_{\rm crit}$ boundary, we select as our sample the high-mass XRB Cygnus X-1 and the low-mass transient XRBs GX 339-4 and XTE J1118+480. The basic properties of these three sources are summarised in Table \ref{tbl-sample}.

Cygnus X-1 is amongst the best studied XRBs \citep[e.g.][]{Zdziarski2002,Wilms2006}, and is both bright and persistent, with a luminosity that varies only by a factor or $\simnot5$. Accretion rates are typically around a few per cent of the Eddington limit \citep{DiSalvo2001}, which places this source near to, or slightly above, $\dot{m}_{\rm crit}$.

Also included in our sample is the low-mass XRB GX 339-4, which is a transient source that, unlike Cygnus X-1, varies in accretion rate by several orders of magnitude \citep{Zdziarski2004}. Although its greater distance and lower black hole mass make GX 339-4 less suitable than Cygnus X-1 for short time-scale study with \textit{RXTE}, it does however have clearer state transitions and a more definite outburst cycle.

XTE J1118+480 was included in the sample of \cite{Wu2008}, who monitored the changes in $\Gamma$ during its outburst in 2000 and showed that this source exhibits the typical harder-when-brighter behaviour expected of an XRB accreting at below $\dot{m}_{\rm crit}$. The X-ray spectrum of XTE J1118+480 is characterised by a power-law component with a weak iron line at $\simnot$6.4 keV \citep{Hynes2000,McClintock2001}.

\begin{table*}
	\caption{A summary of the sample.}
	\label{tbl-sample}
	\begin{center}
	\begin{tabular}{@{}cccccc}
		\hline
		Source & LMXB/HMXB$\,^{a}$ & Distance & $M_{BH}$ & RA$\,^{b}$ & Dec$\,^{b}$ \\
		& & [kpc] & [$M_{\varodot}$] \\
		\hline
		Cygnus X-1 & HMXB & 1.86$^{+0.12}_{-0.11}\,^{c}$ & 14.8$\pm1.0\,^{d}$ & 19h58m21.7s & +35d12m05.8s \\
		GX 339-4 & LMXB & $>7.00\,^{e}$ & 5.8$\,^{f}$ & 17h02m49.4s & -48d47m22.8s \\
		XTE J1118+480 & LMXB & 1.81$\pm0.24\,^{g}$ & 6.9 - 8.2$\,^{h}$ & 11h18m10.8s & +48d02m12.6s \\
		\hline
	\end{tabular} \\
	\end{center}
	\textsc{Notes}: $^{a}$ Low-mass XRB or high-mass XRB, $^{b}$ From SIMBAD Astronomical Database, $^{c}$ \cite{Reid2011}, $^{d}$ \cite{Orosz2011}, $^{e}$ \cite{Zdziarski2004}, $^{f}$ \cite{Hynes2003}, $^{g}$ Average of distances estimated by \citealt{Wagner2001} (1.9$\pm$0.4 kpc), \citealt{McClintock2001} (1.8$\pm$0.6 kpc) and \citealt{Gelino2006} (1.72$\pm$0.1 kpc), $^{h}$ \cite{Khargharia2013} \\
\end{table*}

\subsection{Data selection}

All observations used in this paper have been retrieved from the \textit{RXTE} archive, and were identified based upon our requirements that the PCA configuration remains the same for all observations obtained for each source, the data demonstrate a long-term variability in either the count rate (CR) or hardness ratio (HR), and the observations are of hard or intermediate state and are evenly sampled in time over a reasonably large number of observations. For Cygnus X-1 we were restricted by the tendency for the high time-resolution coverage of the 3-20 keV energy range to be split between binned array mode for channels 0 to 35 (which roughly cover the 2-15 keV band) and event mode for channels 36 and above; all observations with PCA configurations split in this way were ignored.

\begin{figure}
        \centering
        \includegraphics[width=90mm]{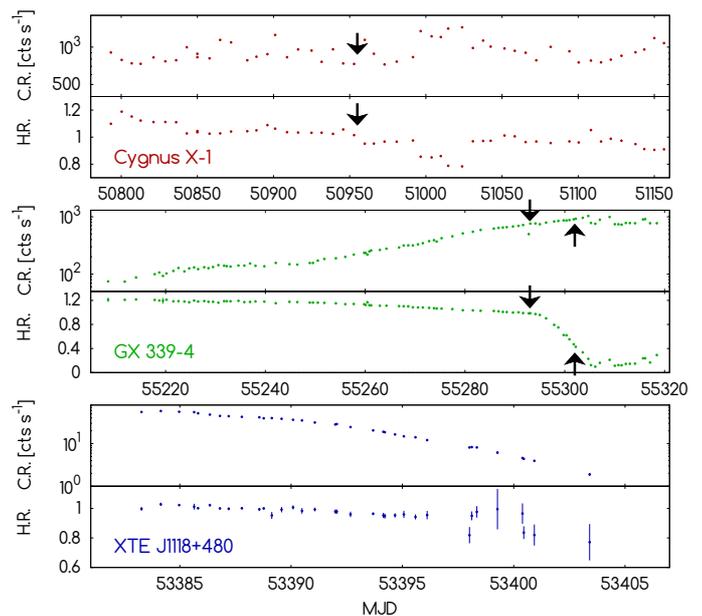}
        \caption{The long-term 3-20 keV PCA (standard-2 mode) count rate (CR) and hardness ratio ($\rm HR  = CR_{6.5-20~keV} / CR_{3-6.5~keV}$; where the 3-6.5 keV and 6.5-20 keV energy ranges have been chosen to be the soft and hard band respectively) for the selected observations of Cygnus X-1 (1997 December 11 to 1998 December 3), GX 339-4 (2010 January 12 to 2010 May 2) and XTE J1118+480 (2005 January 14 to 2005 February 26). Downward arrows mark points in the lightcurves which we later identify (see Section \ref{sec-rms}) as suspected state transitions from the hard to hard-intermediate state (Cygnus X-1: between 1998 May 19, obs. 24, and 1998 May 26, obs 25; GX 339-4: between 2010 April 2, obs. 63, and 2010 April 4, obs. 64), and upward arrows similarly mark suspected hard-intermediate to soft-intermediate state transitions (GX 339-4: between 2010 April 14, obs. 74, and 2010 April 16, obs. 75).}
        \label{fig-lightcurve}
\end{figure}

We have therefore selected 52 observations of Cygnus X-1, all of which have proposal ID 30157 and date from between 1997 December 11 and 1998 December 3. The observations offer the binned array mode PCA configuration, and provide 42 channels to cover the whole 3-20 keV energy range. The exposure times are of between 1.5 ks and 4.4 ks, which, once good time intervals (GTIs) had been generated (see below), provided usable exposure times of between 0.9 ks and 3.4 ks. The lightcurve of count rate and HR produced from these observations is shown in Fig. \ref{fig-lightcurve} (top). Here, and throughout this paper, we choose to use the count rate in the 3-6.5 keV and 6.5-20 keV energy bands in order to calculate HR (defined as $\rm HR = CR_{6.5-20~keV} / CR_{3-6.5~keV}$); these bands have been selected for two reasons: firstly, they provide HRs which are consistently close to unity (and should therefore produce similar levels of statistical variation in each band) and, secondly, the iron line at $\simnot$6.4 keV is roughly divided between the two bands, thereby reducing the effect this component may have upon the HRs.

For GX 339-4 we have selected 91 observations from proposal 95409, dating from between 2010 January 12 and 2010 May 2. These observations all offer the event mode PCA configuration, and provide 25 channels to cover the 3-20 keV energy range. Once GTIs had been applied then the usable exposure times were between 750 s and 3.3 ks, with two additional observations offering longer exposures of 4.5 ks and 9.3 ks; 20 observations with good time intervals of less than 750 s were ignored, as were all soft-state observations which followed after 2010 May 2. The lightcurve of count rate and HR is shown in Fig. \ref{fig-lightcurve} (centre).

Finally, for XTE J1118+480 we selected 30 observations from proposals 90011 and 90111, dating from between 2005 January 14 and 2005 January 28. These observations offer the good xenon mode PCA configuration, and provide 42 channels to cover the 3-20 keV energy range. The GTIs have usable exposure times of up to 4.7 ks, although eight observations were found to be less than 750 s in length and were ignored. The lightcurve of count rate and HR is shown in Fig. \ref{fig-lightcurve} (bottom).

A summary of the observations used in this paper is provided in Table \ref{tbl-observations}.

\begin{table}
	\caption{A summary of the observations used in this paper. Only the first and last observation in each range of observations are shown, and all those in between (indicated with $\hdots$) are numbered in order of observation ID, rather than date.}
	\label{tbl-observations}
	\begin{center}
	\begin{tabular}{@{}cccc}
		\hline
		Source & Obs. & \textit{RXTE} Obs. ID & Date \\
		& Num. & & \\
		\hline
		Cygnus X-1 & 1 & 30157-01-01-00 & 1997 Dec. 11 \\
		& $\hdots$ & $\hdots$ & $\hdots$ \\
		& 52 & 30157-01-51-00 & 1998 Dec. 3 \\[1.5ex]
		GX 339-4 & 1 & 95409-01-01-00 & 2010 Jan. 12 \\
		& $\hdots$ & $\hdots$ & $\hdots$ \\
		& 91 & 95409-01-17-02 & 2010 May 2 \\[1.5ex]
		XTE J1118+480 & 1 & 90011-01-01-00 & 2005 Jan. 14 \\
		& $\hdots$ & $\hdots$ & $\hdots$ \\
		& 14 & 90011-01-01-13 & 2005 Jan. 20 \\
		& 15 & 90111-01-01-00 & 2005 Jan. 13 \\
		& $\hdots$ & $\hdots$ & $\hdots$ \\
		& 45 & 90111-01-07-00 & 2005 Feb. 26 \\
		\hline
	\end{tabular} \\
	\end{center}
\end{table}

\subsection{Data reduction}

In total, 173 observations were retrieved from the \textit{RXTE} archive, and GTIs were generated based upon the spacecraft elevation angle being greater than $10^{\circ}$, the pointing offset being less than $0.02^{\circ}$ and the time since the last south Atlantic anomaly (SAA) passage being at least 30 minutes. Next, synthetic background data based upon the bright PCA background models were generated using the script \textsc{runpcabackest}, and suitable response files created using the \textsc{pcarsp} tool.

For each observation, the data reduction was performed upon both the standard-2 datasets and the higher time resolution binned or event mode data. The standard-2 data offer excellent spectral resolution, but the time resolution of $\rm16~s$ is insufficient for some of the later analysis in this paper. Spectra were extracted from both the standard-2 data and background data for all layers of PCU 2 using the \textsc{saextrct} tool, and these data were used to construct the source lightcurves shown in Fig. \ref{fig-lightcurve}.

The method of reducing high time-resolution PCA data is dependent upon the mode in which the data are supplied, and it was therefore necessary to follow a slightly different process for each of our three sources. For XTE J1118+480 we first used the \textsc{make$\_$se} script to generate event mode data from the good xenon data, from which spectra could subsequently be extracted using the \textsc{seextrct} tool. The data from the GX 339-4 observations were already provided in event mode format, and could be extracted with \textsc{seextrct} immediately. For both XTE J1118+480 and GX 339-4 the spectra were extracted from all layers of PCU 2 only. For Cygnus X-1, spectra were extracted from the binned array mode data using the \textsc{saextrct} tool. However, for this source we were restricted by the less flexible format of the binned array mode (in which counts from all detectors are binned into a single column) to using all five PCUs. The response files for the event and array mode data were generated using the \textsc{pcarsp} tool, and background spectra created by rebinning the background files generated from the standard-2 data.

\section{Spectral analysis of complete observations}
\label{sec-spectral-fitting}

\subsection{Definitions of time-scales used in this paper}

Before we look for spectral variations within individual observations we first analyse the standard-2 mode spectra that were extracted from each complete observation (typically 1 - 3 ks in length), which we shall hereafter refer to as the 'observation-averaged' spectra. Similarly, we shall hereafter refer to any study of the source behaviour across many separate observations (for which we only use the standard-2 data) as being on long time-scales (days to weeks), and any study of the source behaviour within a single observation (for which we only use the higher time-resolution modes) as being on short time-scales (seconds to minutes). Furthermore, the term 'observation' shall only be used in reference to the entire \textit{RXTE} observation, and never to describe any short time-scale section of data.

\subsection{Hardness-intensity diagrams}

\begin{figure*}
        \centering
        \includegraphics[width=150mm]{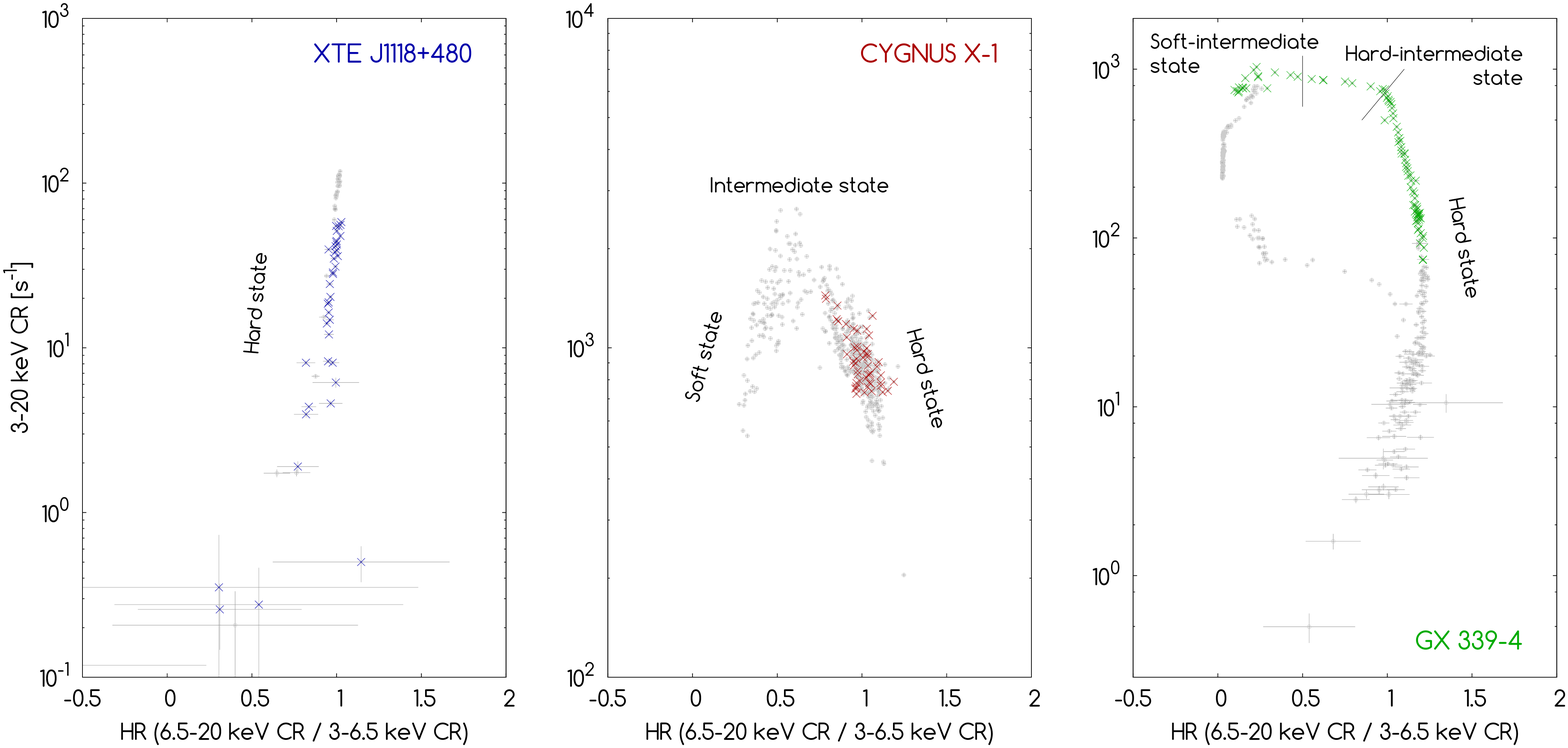}
        \caption{Long time-scale hardness-intensity diagrams (HR in the 3-6.5 keV and 6.5-20 keV bands versus 3-20 keV count rate) for XTE J1118+480, Cygnus X-1 and GX 339-4, extending over the periods 2000 April 13 to 2005 January 28, 1998 December 25 to 2004 February 20 and 2007 June 21 to 2011 April 5 respectively. The HRs are based upon spectra extracted from the standard-2 mode data. The coloured crosses indicate the observations that have been selected for further analysis in this paper, and are all of hard or intermediate spectral state.}
        \label{fig-hr-long-term}
\end{figure*}

Fig. \ref{fig-hr-long-term} shows the long time-scale hardness-intensity diagrams (HIDs) for the three sources, based upon the background subtracted count rates of the standard-2 data. Each data point represents one complete observation (which are typically a few kiloseconds in length), and observations are separated in time by intervals of between a few hours and several days. The data from the observations selected for analysis in this paper (denoted with coloured crosses) are shown alongside other observations from around the same time in order to provide a more complete picture of the evolution of each source on these plots.

The HID of GX 339-4 takes the form of the familiar q-shaped hysteresis curve (sometimes called the turtle-head diagram), the head of which is always traversed in an anti-clockwise direction \citep{Fender2004,Belloni2005}. The observations selected for analysis in this paper begin in the hard state at the point at which the source switches from harder-when-brighter to softer-when-brighter behaviour, and continue through the intermediate state (the top horizontal branch of the distribution). The fainter hard-state observations, in which the source is harder when brighter, were found to be too faint to examine on 100 ms time-scales. In contrast, Cygnus X-1 does not exhibit clear state transitions between the hard and intermediate states in its HID (although the soft state can be seen towards the left-hand side of the plot), so we do not yet attempt to distinguish between the two; it is clear, however, that none of these data is expected to be of the soft state. Finally, XTE J1118+480 is at a low accretion rate and is always found in the hard state. The HID of this source shows only harder-when-brighter behaviour, and is therefore comparable to the tail of the "q" in the HID of GX 339-4.

The absence of hysteresis in the HID of Cygnus X-1 has been attributed \cite[e.g.][]{Smith2002,Maccarone2003b} to the high-mass nature of the donor and the wind-fed accretion mechanism, which are expected to form smaller accretion discs than those found in low-mass transient systems. \cite{Belloni2010} further adds that the weaker the outburst, the smaller the difference in flux between the hard-to-soft and the soft-to-hard state transitions, and that state transitions in Cygnus X-1 represent hysteresis that is too small to observe.

\subsection{Single power-law fits}

The spectra extracted from the standard-2 mode data are initially fitted in \textsc{xspec} version 12.8 using just a single power-law model (which, at this stage, does not include absorption). Whereas a model of such simplicity cannot adequately subsume other known features such as the iron line or accretion disc, the residuals to the fit do however provide a useful first look at the spectral components that are present, and the model therefore provides a suitable basis upon which further complexity can be added if required.

\begin{figure*}
        \centering
        \includegraphics[width=150mm]{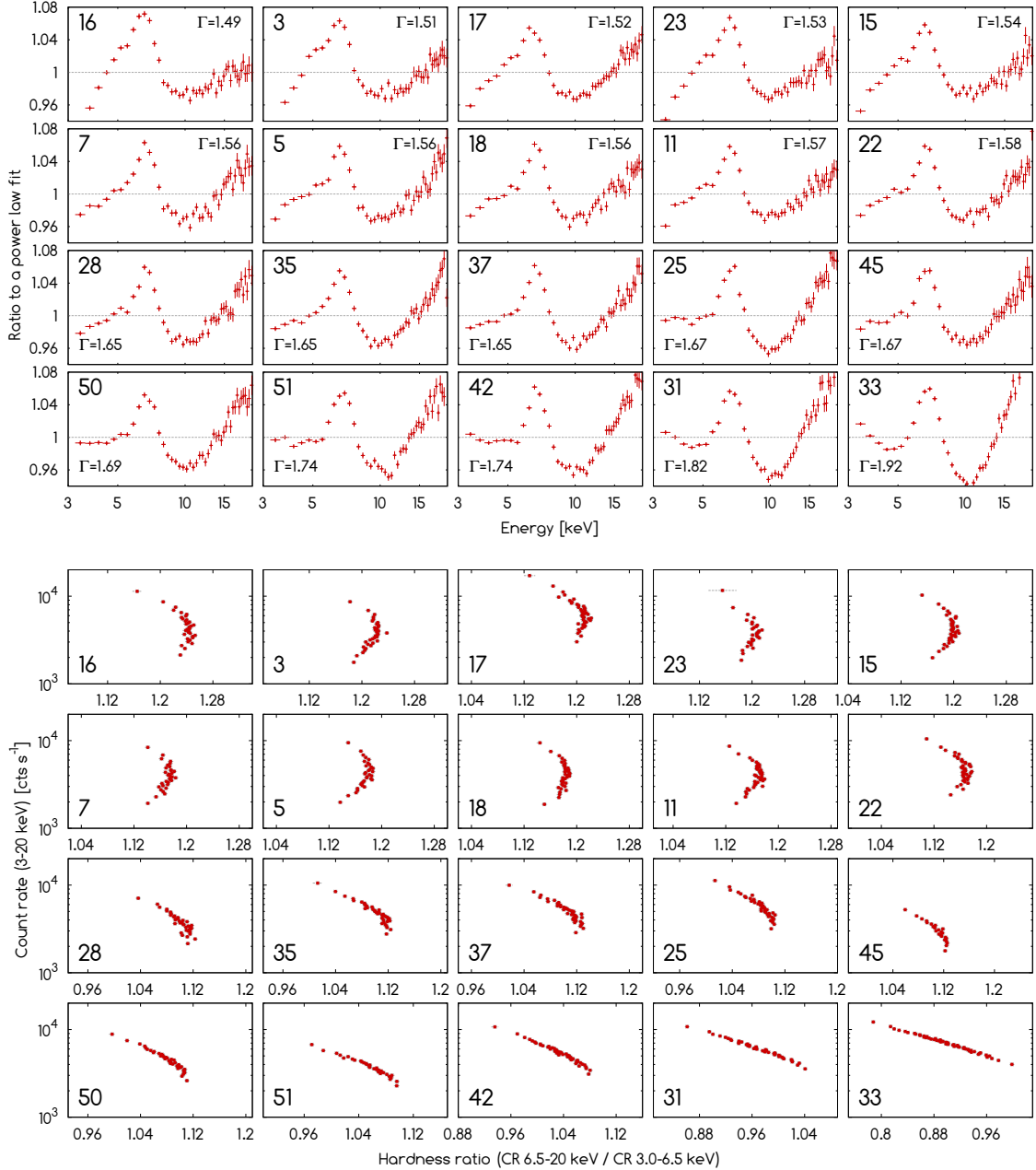}
        \caption{\textit{\bf Top:} The standard-2 mode PCA energy spectra from 20 of the Cygnus X-1 observations, sorted by decreasing HR (from left to right on the top row, then the second row, etc). The spectra are shown as a ratio to a power-law fit, and the photon index is shown in each panel. Spectra with large $\Gamma$ often show a change in spectral slope at around 10 keV. In Section \ref{sec-rms} we tentatively associate obs. 1-24 (the top two rows) with the hard state, and obs. 25-52 (the bottom two rows) with the hard-intermediate state. \textit{\bf Bottom:} Short time-scale hardness-intensity diagrams for the same 20 Cygnus X-1 observations. The count rates shown are all background subtracted. Observations with low $\Gamma$ tend to show harder-when-brighter behaviour at low count rates, and softer-when-brighter behaviour at high count rates. In contrast, observations with high $\Gamma$ always show a very strong softer-when-brighter behaviour at all times.}
        \label{fig-hr-cyg-short-term}
\end{figure*}

Fig. \ref{fig-hr-cyg-short-term} (top) shows the standard-2 mode spectra from 20 observations of Cygnus X-1, which have been sorted by decreasing HR and are presented in the form of ratios to the power-law fits. Each of these spectra show a clear excess at around 6 or $\rm7~keV$, consistent with a broad iron line. In addition to the iron line, the spectra also reveal the presence of at least one further component which is responsible for causing a change in spectral slope at approximately 10 keV in some of the brighter, softer observations (see, for example, obs. 31 and 33). This additional component could either be attributed to a combination of thermal disc emission and/or reflection, both of which one would expect to increase in strength in the softer, brighter observations due to the inward motion of the disc, or otherwise to an additional Comptonised power-law component.

\subsection{Adding an additional continuum component: disc or power-law?}
\label{sec-fitting-additional-component}

The process of decomposing these spectra into their separate components is made more complicated by absorption from a large ($N_{\rm H} > 10^{21}~{\rm cm}^{-2}$) column density. Furthermore, if two separate power-laws are present then it is not possible to disentangle them from their associated reflection components due to the high level of degeneracy in the fit. Therefore, the Cygnus X-1 spectra were fitted with two different models: the first (hereafter referred to as the PL+PL model) consisted of two power-laws (\textsc{powerlaw}), a Gaussian representing the iron line (\textsc{gaussian}) and photoelectric absorption (\textsc{phabs}, with $N_{\rm H}$ approximately $6 \times 10^{21}~{\rm cm}^{-2}$ in most of the observations and around 2-4 times higher in the remaining few; the absorption was applied to both power-laws), and the second (hereafter referred to as the PL+R model) consisted of a single power-law plus reflection (\textsc{pexrav}, see \citealt{Magdziarz1995}), an iron line and photoelectric absorption. Whilst fitting the reflection component, the inclination angle of the disc was fixed to $27.1^{\circ}$ \citep{Orosz2011}.

Physically, we interpret the PL+R model as representing a corona and reflection from an accretion disc, and the PL+PL model as representing the two-component Comptonised emission arising from the inner and outer accretion flows. We do not initially include a reflection component in our PL+PL model due to concerns over degeneracy in our fits, although we do expect there to be some reflection from the truncated accretion disc.

\begin{figure}
	\centering
        \includegraphics[width=70mm]{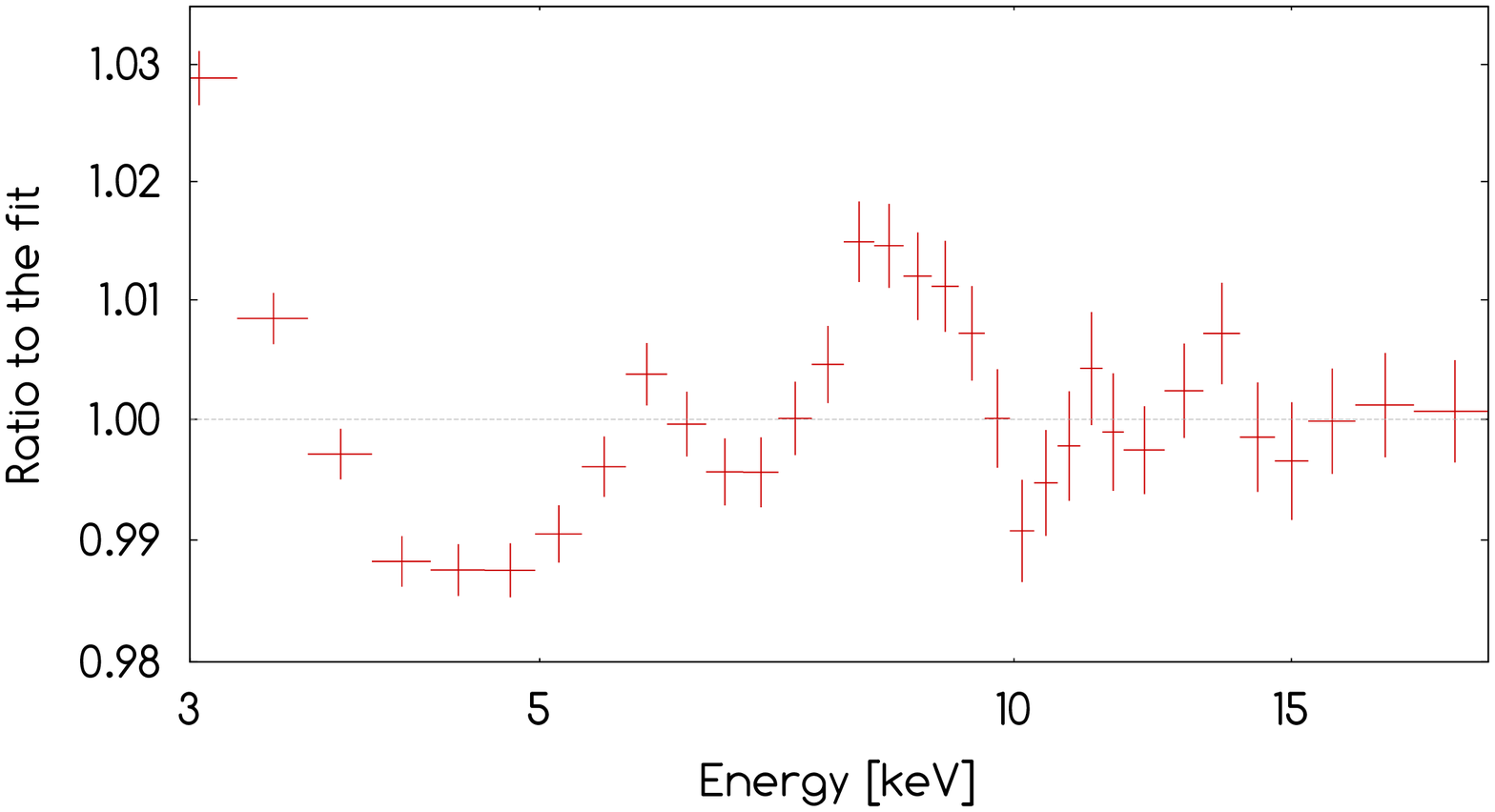}
        \caption{An example (from obs. 33) of a typical soft excess found below 5 keV in the spectra of the brighter, softer Cygnus X-1 observations. The energy spectrum was fitted with a model consisting of an absorbed power-law, an iron line and reflection (referred to in this paper as the PL+R model), and the plot shows the ratio to this fit. We find that adjusting the column density of the absorption fails to improve the fit at low energies, and we instead attempt to fit this excess emission with either an accretion disc component or an additional power-law.}
        \label{fig-cyg-soft-excess}
\end{figure}

When the data were fitted with the PL+R model we found a steep soft excess below $\rm\simnot4$-5 keV in approximately one half of the observations (see Fig. \ref{fig-cyg-soft-excess}). When this excess was fitted with an accretion disc (\textsc{diskbb}) the disc normalisation ($K$) failed to converge to a value consistent with a truncation radius greater than one gravitational radius ($R_{\rm G}$) in any of the fits, and $K$ was therefore fixed to $5 \times 10^{4}$ (a truncation radius of $\rm\simnot45~km$, or $\simnot2~R_{\rm G}$). It was found that varying $K$ between $2 \times 10^{4}$ and $2 \times 10^{6}$ (a truncation radius of $\rm\simnot28 \mhyphen 280~km$, or $\simnot1 \mhyphen 12~R_{\rm G}$) made little difference to the quality of the fits (although $\chi^{2}$ tended to increase as $K$ increased), or to the parameters of the power-law, and affected only the inner disc temperature. Increasing $K$ above this range produced a noticeably poorer fit, which is not consistent with the expected large truncation radius of the disc in the hard or hard-intermediate state.

\begin{figure}
	\centering
        \includegraphics[width=86mm]{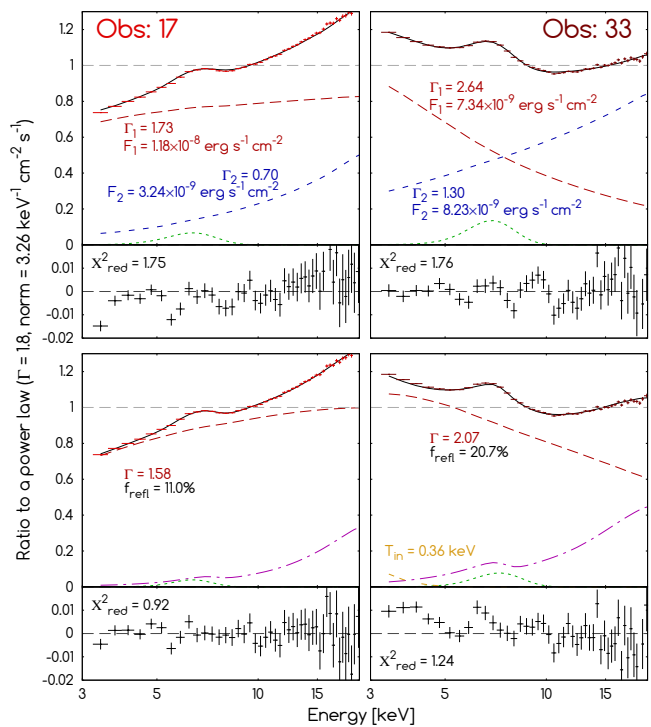}
        \caption{\textit{\bf Top:} spectral fits (solid black lines) with residuals (smaller panels) to Cygnus X-1 obs. 17 and 33 using the PL+PL model, which consists of two power-laws (long-dashed red and short-dashed blue lines), a Gaussian representing the iron line (dotted green line) and photoelectric absorption. The spectra are shown as a ratio to a single power-law ($\Gamma = 1.8$). \textit{\bf Bottom:} spectral fits (solid black lines) with residuals (smaller panels) to the same data using the PL+R model, which consists of a single power-law (long-dashed red line), reflection (dot-dashed magenta line), an iron line (dotted green line) and an accretion disc (double-dashed orange line).}
        \label{fig-cyg-17-33-components}
\end{figure}

Fig. \ref{fig-cyg-17-33-components} shows two observations of Cygnus X-1 (17 and 33), which were selected because they have similar count rates but very different HRs; in addition, obs. 33 shows a clear change in spectral slope at $\simnot$10 keV, whereas obs. 17 does not. The parameters derived from the PL+R model are not unreasonable, with reflected fractions of 11.0 per cent and 20.7 per cent in obs. 17 and 33 respectively. In addition, the soft excess in obs. 33 was fitted with an accretion disc with a temperature at the inner disc radius of 0.358 keV. A summary of the fit parameters for obs. 17 and 33 is provided in Table \ref{tbl-fit-parameters}.

After fitting all 52 Cygnus X-1 observations with the same two models, it was found that, in general, the PL+R model offered a slightly better fit than the PL+PL model, but both models consistently delivered a reduced $\chi^{2}$ ($\chi^{2}_{\rm red}$) of between 0.8 and 2, and it is not possible with these data to determine which model is the most appropriate. Whilst fitting obs. 33 we found that the fit from the PL+PL model could be improved considerably ($\chi^{2}_{\rm red} = 0.92$) by including a component to model the reflection from the soft power-law, but any attempt to expand the model to include reflection from the hard power-law resulted in degeneracy between the hard power-law photon index ($\Gamma_{2}$) and the reflection scaling factor, and the $\chi^{2}_{\rm red}$ did not improve.

\begin{table}
	\caption{A summary of the fit parameters for obs. 17 and 33, fitted with two different models. In addition to the components mentioned below, both models also included an iron line at $\rm\simnot6.4~keV$ and photoelectric absorption ($N_{\rm H} = 6 \times 10^{21}~{\rm cm}^{-2}$).}
	\label{tbl-fit-parameters}
	\begin{center}
	\begin{tabular}{|p{5cm}|c|c|}%{@{}lcc}
		\hline
		Model/ & Obs 17 & Obs 33 \\
		Parameter \\
		\hline
		\multicolumn{3}{l}{\textsc{powerlaw} + \textsc{powerlaw} (PL+PL model)} \\
		$\chi^{2}_{\rm red}$ (40 d.o.f.)\dotfill & 1.75 & 1.76 \\
		$\Gamma_{1}$\dotfill & 1.73 & 2.64 \\
		$\Gamma_{2}$\dotfill & 0.70 & 1.30 \\[1.5ex]
		\textsc{pexrav} + \textsc{diskbb} (PL+R model) \\
		$\chi^{2}_{\rm red}$ (41 and 44 d.o.f. respectively)\dotfill & 0.92 & 1.24 \\
		$\Gamma$\dotfill & 1.58 & 2.07 \\
		$kT_{\rm in}$ [keV]$\,^{a}$\dotfill & / & 0.358 \\
		$rel_{\rm refl}\,^{b}$\dotfill & 0.46 & 1.13 \\
		$f_{\rm refl}$ [$\%$]$\,^{c}$\dotfill & 11.0 & 20.7 \\
		\hline
	\end{tabular} \\
	\end{center}
	\textsc{Notes}: $^{a}$ Disc temperature at inner radius. The normalisation was fixed to $5 \times 10^{4}$, which is equivalent to an inner radius of $\rm\simnot45~km$, $^{b}$ \textsc{pexrav} reflection scaling factor, $^{c}$ reflection flux divided by power-law flux \\
\end{table}

\section{Short time-scale spectral variability}

\subsection{Overview}

In order to examine spectral variability on short time-scales the high time-resolution data were split into 100 ms segments, and spectra were extracted from each segment using either \textsc{saextrct} or \textsc{seextrct}, as appropriate. On such short time-scales spectra may be severely affected by Poisson noise due to the low count statistics \citep{Wu2010}, so the individual spectra were sorted by count rate and then added together using the \textsc{addspec} tool such that each summed spectrum contains a similar number of background subtracted counts; for Cygnus X-1 this total was chosen to be $300\,000$ counts, for GX 339-4 it was $55\,000$ counts and for XTE J1118+480 it was $21\,000$ counts. The total number of summed spectra for each observation therefore depends upon both the brightness and length of the observation, and we find that we have between 12 and 61 spectra for each Cygnus X-1 observation, between 5 and 35 spectra for each GX 339-4 observation and between 5 and 9 spectra for each observation of the much fainter XTE J1118+480.

\subsection{Short time-scale HIDs}

The short time-scale spectral variability is examined by constructing HIDs from the summed 100 ms spectra, where each data point represents a measurement of count rate and HR from a single summed spectrum, and a separate HID is constructed for each observation.

Fig. \ref{fig-hr-cyg-short-term} (bottom) shows the HIDs from 20 observations of Cygnus X-1, which are sorted by decreasing HR. Many of the panels show a weak softer-when-brighter behaviour, which becomes much stronger in the observations with the softest spectra (e.g. obs. 31 and 33). Conversely, some observations (e.g. obs 7, 11, 15, etc) have slightly crescent-shaped distributions and show signs of harder-when-brighter behaviour at low count rates. In general, crescent-shaped HIDs are found only when $\rm HR > 1.12$ (roughly equivalent to $\Gamma < 1.6$). Although the shape of the short time-scale HIDs is mostly dependent upon HR, the weak correlation between count rate and HR means that many of the strong softer-when-brighter HIDs are found in the brightest observations.

The data also suggest that the shape of the HID evolves fairly slowly in time, and observations only a few weeks apart tend to have similar HIDs even when the count rate has changed noticeably (for example, see obs. 25 and 28, and obs. 16 and 17). The long time-scale evolution in HR and shape of the HID (for instance, the disappearance of the crescent-shaped HIDs and a decrease in HR to a value less than 1.12) coincide with the appearance in the energy spectra of the additional soft component below 10 keV, the origin of which will be further examined in the next section.

\begin{figure*}
        \centering
        \includegraphics[width=150mm]{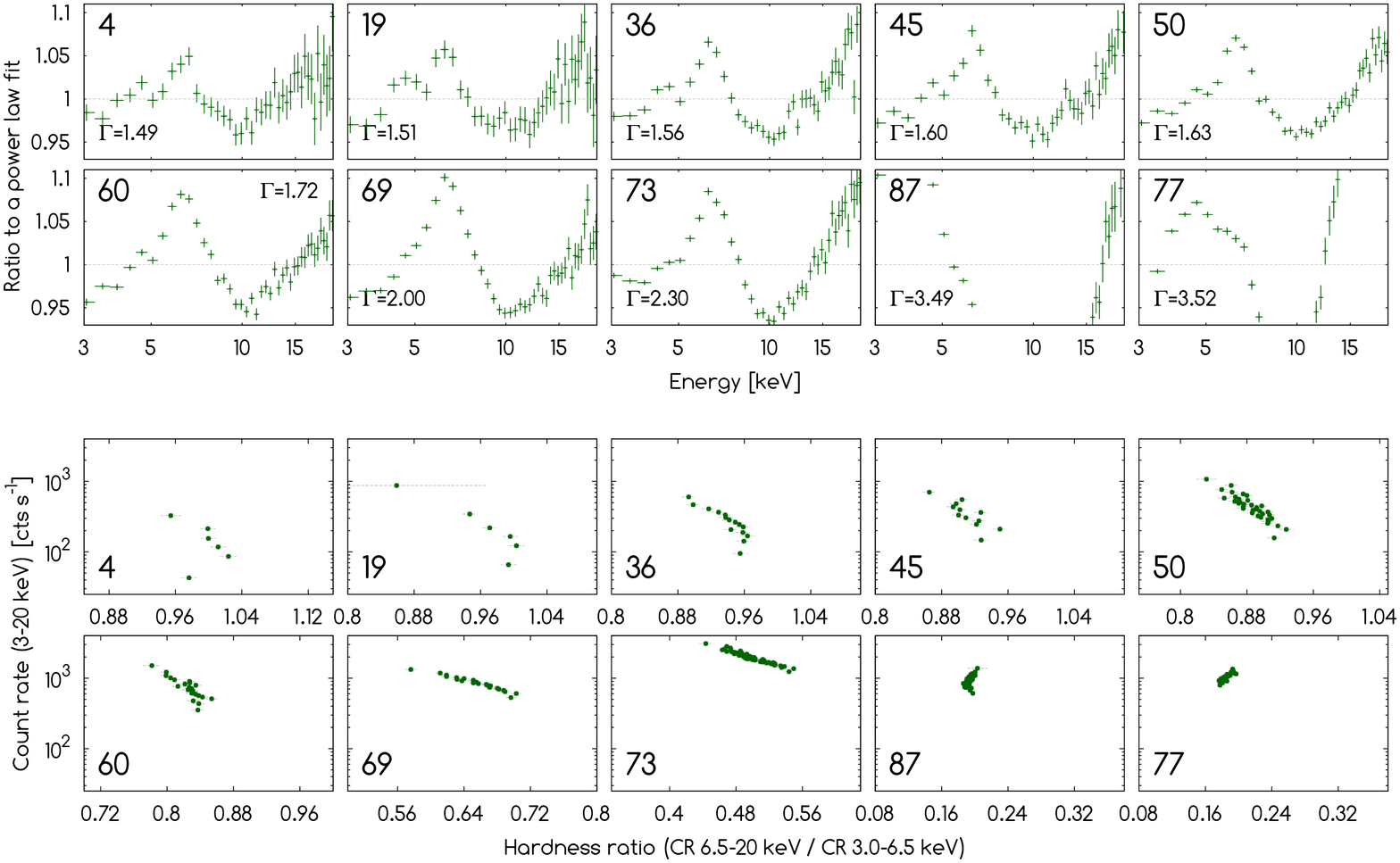}
        \caption{\textit{\bf Top:} The energy spectra from ten of the GX 339-4 observations, sorted by decreasing HR, and shown as a ratio to a power-law fit. As the count rate increases the spectra become gradually softer, before eventually becoming dominated by thermal emission in the soft-intermediate state (obs. 75+). The increase in the equivalent width of the iron line whilst in the hard and hard-intermediate state is also clear. \textit{\bf Bottom:} Short time-scale hardness-intensity diagrams for the same ten observations of GX 339-4. Most of the hard-state observations (obs. 1-63) show a softer-when-brighter behaviour, which becomes stronger once the source makes the transition to the hard-intermediate state (obs. 64-74). In the soft-intermediate state (obs. 75+) the growing contribution to the spectra of a relatively stable thermal disc component reduces the variability in HR.}
        \label{fig-hr-gx339-short-term}
\end{figure*}

In general, the short time-scale HIDs of GX 339-4 (Fig. \ref{fig-hr-gx339-short-term}, bottom) show a softer-when-brighter behaviour in the hard state, which is consistent with the behaviour found in the same observations on time-scales of weeks (Fig. \ref{fig-hr-long-term}, right). Once the source enters the hard-intermediate state (obs. 64+) then the spectra become softer, and the softer-when-brighter behaviour appears stronger. At a later time still (roughly obs. 75+) there is a transition from a hard-intermediate to a soft-intermediate state, with a reduction in the spectral variability of the source.

\begin{figure*}
        \centering
        \includegraphics[width=150mm]{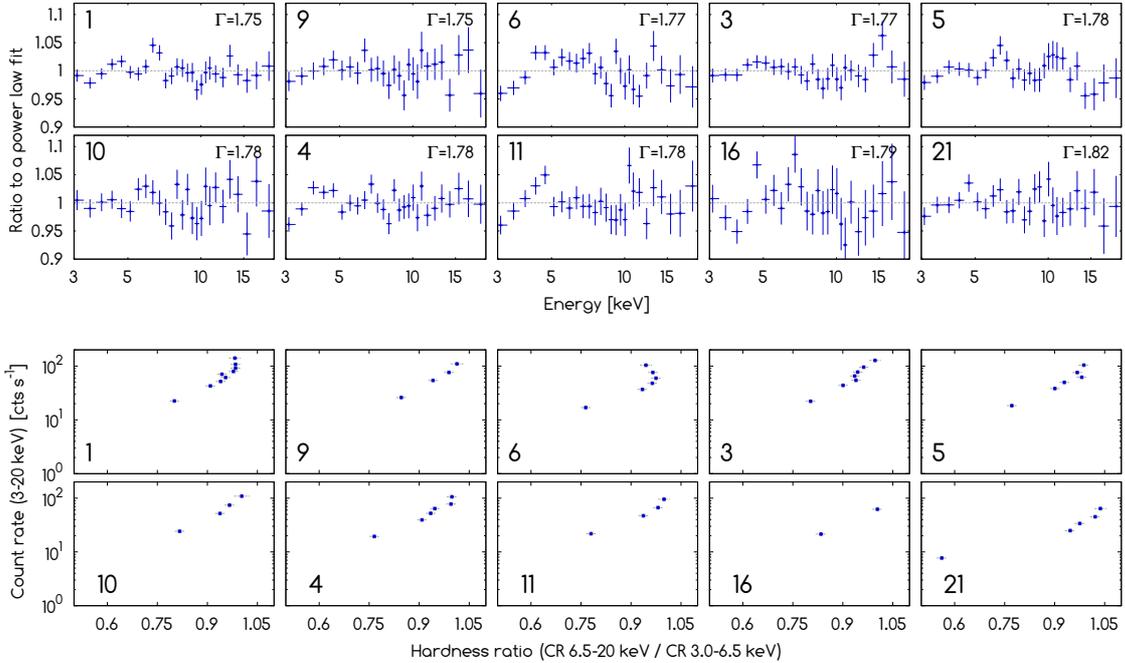}
        \caption{\textit{\bf Top:} The energy spectra from ten of the brightest XTE J1118+480 observations, sorted by decreasing HR, and shown as a ratio to a power-law fit. All the spectra appear consistent with a single power-law fit, with only a few of the observations showing any indication of an iron line. \textit{\bf Bottom:} Short time-scale hardness-intensity diagrams for the same ten observations of XTE J1118+480. All observations show a harder-when-brighter behaviour, which is consistent with Comptonisation in a radiatively inefficient accretion flow.}
        \label{fig-hr-j1118-short-term}
\end{figure*}

For XTE J1118+480 the short time-scale HIDs (Fig. \ref{fig-hr-j1118-short-term}, bottom) show only the typical harder-when-brighter behaviour expected from a low accretion rate source, and are therefore similar to the HID produced on time-scales of days (Fig. \ref{fig-hr-long-term}, left). Since this source does not undergo any state transitions, and its accretion rate does not increase above $\dot{m}_{\rm crit}$, then the short time-scale HIDs show few changes between observations.

\subsection{Spectral fitting of short time-scale data}

In Section \ref{sec-spectral-fitting} we found that the Cygnus X-1 observation-averaged spectra could be fitted equally well with a dual power-law model (PL+PL) or with a single power-law plus reflection model (PL+R, including an accretion disc if required). Here, we fit the short time-scale, flux-binned spectra with the same two models in order to compare the quality of fit, and also determine how the spectral index varies as a function of the Eddington ratio ($L_{\rm3-20~keV}/L_{\rm edd}$, hereafter denoted $L_{\rm X/E}$) on short time-scales.

Whilst fitting the short time-scale data we choose to keep some of the parameters, such as the disc temperature and iron line equivalent width (EW), fixed to the values that were fitted to the observation-averaged spectra (see \citealt{Skipper2013} for a discussion of the validity of fixing these parameters). In order to fix the iron line EW, a \textsc{gabs} component with a negative line depth was used in place of the Gaussian emission component.

We find that both the PL+PL and PL+R models are successfully able to fit the short time-scale spectra of Cygnus X-1, with an average $\chi^{2}_{\rm red}$ of between 1 and 1.2 for the 133 flux-binned spectra extracted from obs. 5, 33 and 37 (where $0.5 < \chi^{2}_{\rm red} < 2.1$ over all 133 spectral fits, with 40 degrees of freedom for both models). For the PL+PL model it is necessary to keep the parameters of the soft power-law component, along with the normalisation of the hard power-law, as free parameters, but the photon index of the hard power-law component can be fixed without any increase in $\chi^{2}_{\rm red}$. For the PL+R model, the photon index, normalisation and reflection scaling factor ($rel_{\rm refl}$) were initially allowed to vary, and it was found that $rel_{\rm refl}$ tends to increase as the flux increases (from 1.02 to 1.32 in obs. 33, equivalent to an increase in reflected fraction from $20.55$ per cent to $21.67$ per cent). However, fixing $rel_{\rm refl}$ made little difference to the average $\chi^{2}_{\rm red}$ (which increased from 1.17 to 1.19 in obs. 33, and from 1.16 to 1.17 in obs. 37).

\begin{figure*}
        \centering
        \includegraphics[width=140mm]{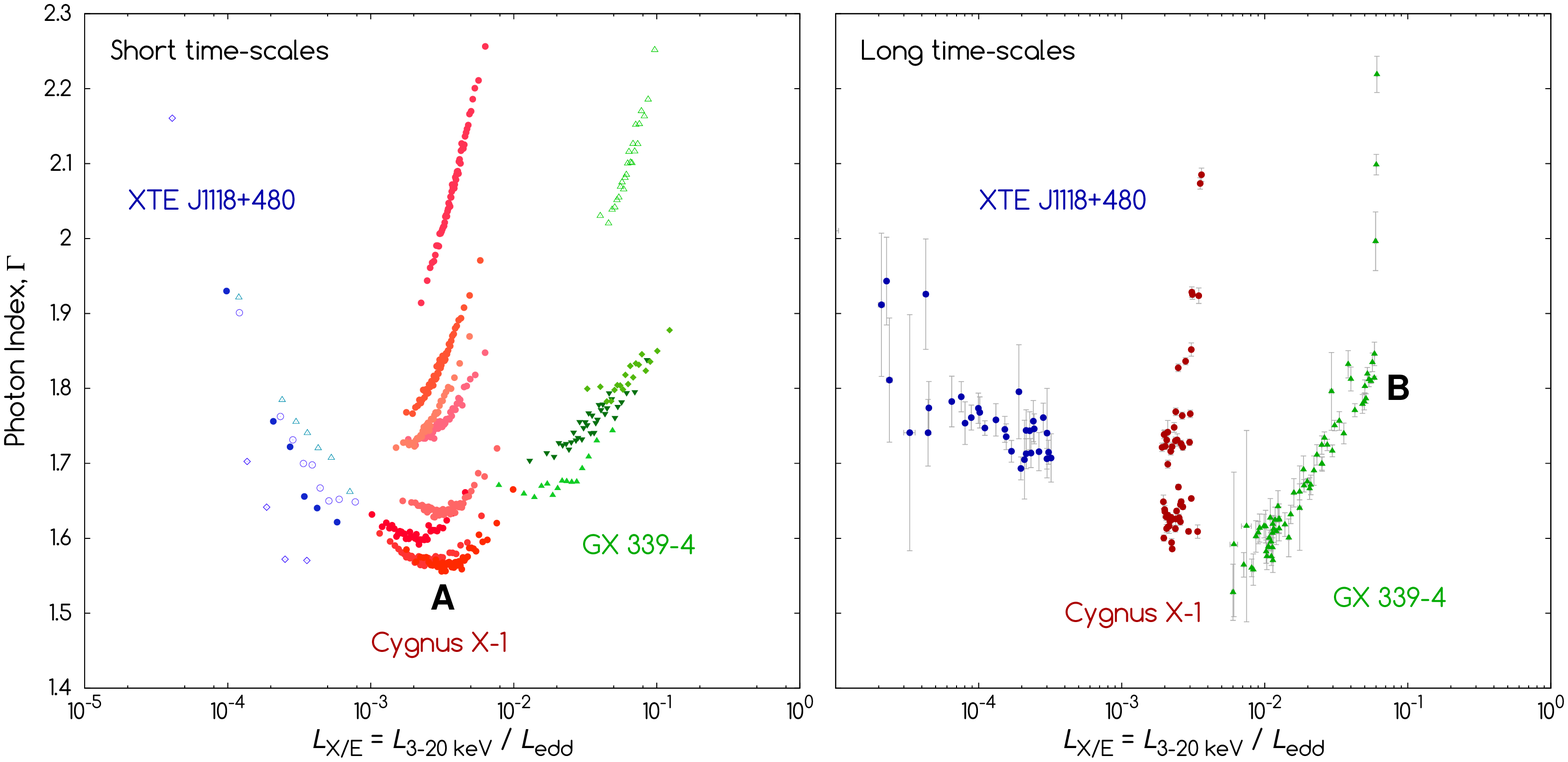}
        \caption{\textit{\bf Left:} short time-scale distribution of $L_{\rm X/E} = L_{\rm3-20~keV}/L_{\rm edd}$ versus $\Gamma$ for a selection of observations. A different symbol and/or colour has been applied to data extracted from each observation. The spectra were fitted with the PL+R model. \textit{\bf Right:} long time-scale distribution of $L_{\rm X/E}$ versus $\Gamma$ for the hard-state observations, based upon spectral fits to the standard-2 PCA data. The Cygnus X-1 data show the evolution of the source over a period of one year, whereas the GX 339-4 data extend over a period of $\simnot3$ months and the XTE J1118+480 data over $\simnot2$ weeks. For XTE J1118+480 and GX 339-4 we find little difference in behaviour on these two different time-scales, whereas on short time-scales many of the Cygnus X-1 observations switch at the point labelled 'A' from a positive correlation at high count rates to an anti-correlation at low count rates.}
        \label{fig-edd-ratio-v-gamma}
\end{figure*}

Fig. \ref{fig-edd-ratio-v-gamma} (left) shows the short time-scale distribution of $\Gamma$ with $L_{\rm X/E}$ for a selection of observations (each of which is plotted with a different symbol and colour), and is contrasted (Fig. \ref{fig-edd-ratio-v-gamma}, right) with the long time-scale distribution obtained from fits to the standard-2 data from each full observation. All spectra were fitted with the PL+R model (including an accretion disc, if required), with the reflected fraction, iron line EW and disc temperature and normalisation fixed whilst fitting the short time-scale data. The disc inclination angle was fixed to $27.1^{\circ}$ for Cygnus X-1 \citep{Orosz2011}, $50^{\circ}$ for GX 339-4 (the inclination angle is not well constrained, but we use the value suggested by \citealt{Shidatsu2011}) and $73.5^{\circ}$ for XTE J1118+480 (the mid-point of the $68 - 79^{\circ}$ estimate of \cite{Khargharia2013}; although this measurement relates to the binary inclination rather than the disc inclination we have assumed that the disc will roughly follow the binary plane). The X-ray luminosity to Eddington luminosity ratios were estimated from the fluxes by using the masses and distances listed in Table \ref{tbl-sample}. For XTE J1118+480 the absorption column density was fixed to $\rm 1.34 \times 10^{20}~{cm}^{-2}$.

For GX 339-4 and XTE J1118+480 we find little difference in the distribution of $\Gamma$ on long and short time-scales, but for Cygnus X-1 the short time-scale data show a marked deviation from the steep positive correlation found in the long time-scale data. When $\Gamma$ is low, the Cygnus X-1 data show the same crescent-shaped form that was evident in the HIDs, but as $\Gamma$ increases then the slope of the $L_{\rm X/E} - \Gamma$ correlation gradually approaches that found on longer time-scales. The transition from the hard to the hard-intermediate state in GX 339-4 causes the rapid rise in $\Gamma$ at point 'B' in Fig. \ref{fig-edd-ratio-v-gamma} (right).

\section{RMS variability}
\label{sec-rms}

\subsection{Count rate versus total rms}

The linear relationship between the flux (in counts per second) and root mean squared (rms) variability in the hard state of Cygnus X-1 was first revealed by \cite{Uttley2001}, who also suggested that the non-zero intercept of this relation with the flux axis implied the presence of a second, non-varying component to the light curve. The rms-flux relation has since been found in a wide variety of sources, including XRBs, AGN, a white dwarf and an ultraluminous X-ray source \citep[see][and references therein]{Heil2012}.

Here we choose to use the method described by \cite{Arevalo2008} (and references therein), in which the variance $\sigma^{2}_{{\rm N},\Delta f}$ is derived from summing the discrete power density spectrum (PDS) over the required frequency interval $\Delta f$, and then subtracting the contribution from Poisson noise to obtain the normalised excess variance $\sigma^{2}_{{\rm NXS},\Delta f}$. Poisson noise is expected to contribute to the normalised variance by adding a component that is almost constant at all frequencies \cite[see also][]{Vaughan2003}, and here we subtract $2\,\bar{x}_{tot}/\bar{x}^{2}$ (where $\bar{x}_{tot}$ and $\bar{x}$ are the total count rate and background subtracted count rate respectively) from each frequency bin of the PDS. The rms variability is given by the square root of $\sigma^{2}_{{\rm NXS},\Delta f}$, which is multiplied by the mean count rate to give the total (not normalised) rms, $\sigma_{{\rm rms},\Delta f}$.

We have calculated the rms (in the 5 mHz to 5 Hz frequency range) separately for the hard band (6.5-20 keV), soft band (3-6.5 keV), and for the whole 3-20 keV energy range. The frequency range was constrained at the lower end by our minimum observation length of 750 s, and at the upper end by the shortest time-scale on which we choose to measure $\Gamma$ (100 ms).

\begin{figure*}
	\centering
        \includegraphics[width=120mm]{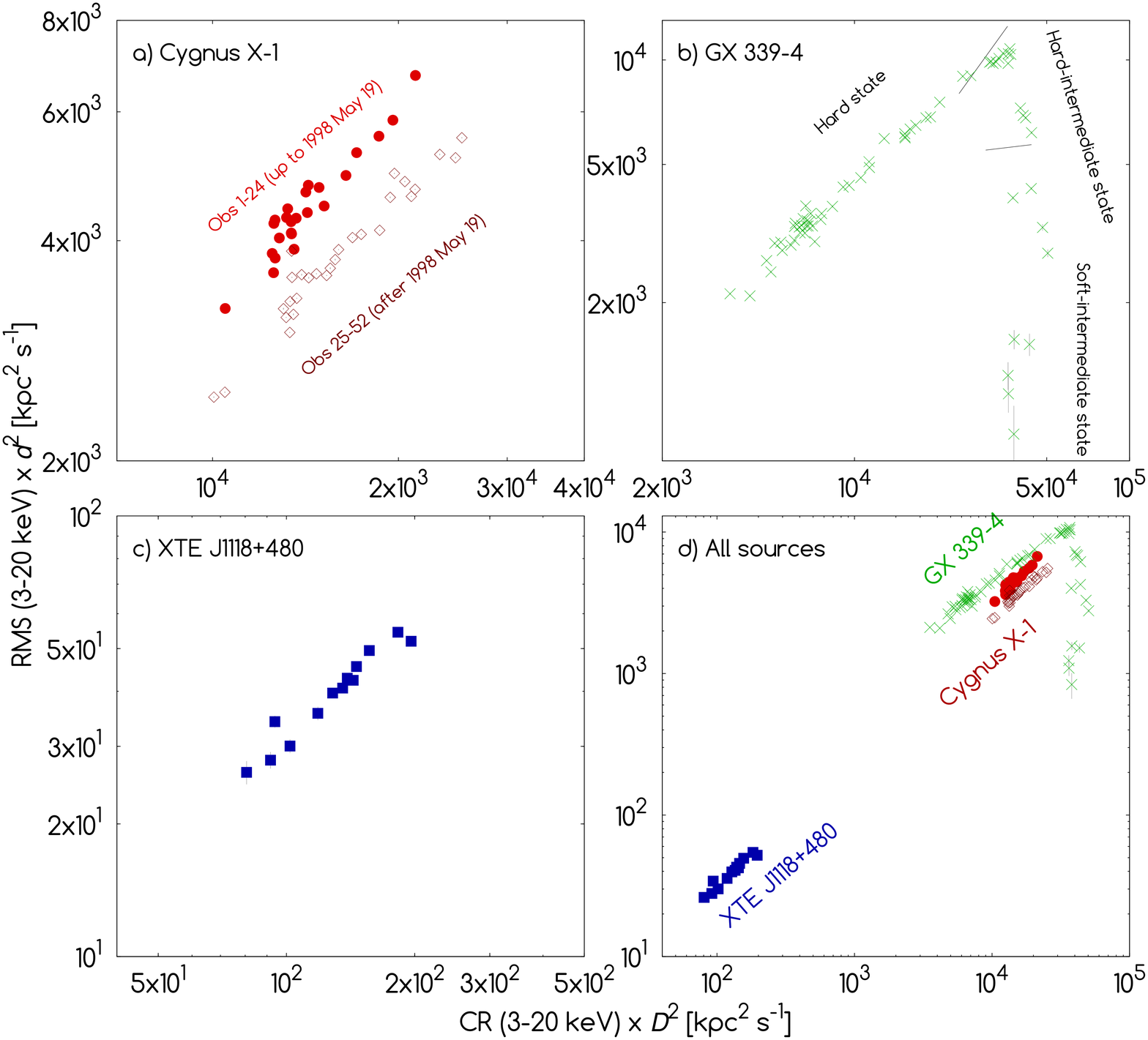}
        \caption{3-20 keV count rate versus rms (in the 0.005-5 Hz frequency range) for Cygnus X-1 (top-left), GX 339-4 (top-right), XTE J1118+480 (bottom-left) and for all three sources (bottom-right). The count rates have all been scaled by the square of the distance ($D$) in order to remove any dependence upon this property, and thereby allow the sources to be directly compared. The GX 339-4 data (green crosses) show a clear distinction between the hard and intermediate states, which are labelled based upon the hardness intensity diagram of GX 339-4 (Fig. \ref{fig-hr-long-term}, right). The Cygnus X-1 data have been divided into observations up to and including 1998 May 19 (obs. 1-24; light red filled circles), and observations after this date (obs. 25-52; dark red open diamonds). The latter data points, which form a separate track on the plot and are characterised by a relatively lower fractional rms and softer spectra, coincide with the change in spectral slope at $\simnot10$ keV in the observation-averaged energy spectra becoming increasingly pronounced (Fig. \ref{fig-hr-cyg-short-term}, top).}
        \label{fig-cr-rms}
\end{figure*}

Fig. \ref{fig-cr-rms} shows the 3-20 keV count rate plotted against the total rms (5 mHz to 5 Hz) for all of our observations. The Cygnus X-1 data consist of one observation every seven days between 1997 December 11 and 1998 December 3, and reveal that the observations from after 1998 May 19 (obs. 25-52) have a lower fractional rms than those from before (obs. 1-24). This reduction in rms has previously been noted by \cite{Pottschmidt2003}, who also found that the power spectra of these data could be well-fitted with four Lorentzians, the peak frequencies of which increased following the reduction in rms; the authors associate some or all of the Lorentzian components with the accretion disc corona. We also note from the observation-averaged energy spectra of Fig. \ref{fig-hr-cyg-short-term} that obs. 25-52 generally have softer spectra, higher count rates, and a more pronounced change in spectral slope at $\simnot10$ keV than obs. 1-24.

Changes in the intercept and/or slope of the rms-flux relation in Cygnus X-1 are often indicative of state transitions or failed state transitions \citep{Gleissner2004}, and we shall hereafter refer to the observations from up to and including 1998 May 19 (obs. 24) as being in the hard state, and those from after this date as being in the hard-intermediate state; similar transitions between these two states have also been observed at other times \cite[e.g.][]{Belloni1999}. For a detailed analysis of state transitions in Cygnus X-1 see \cite{Grinberg2013}.

The GX 339-4 data reveal a marked change in the rms-flux relation during the hard to hard-intermediate state transition that occurred around April 2010 (obs. 63 to 64). The large reduction in fractional rms that follows during the intermediate-state phase is consistent with the results of \cite{Munoz-Darias2011}, who found that the rms-flux relation of GX 339-4 follows a complete hysteresis curve as the source moves through all the states of a typical outburst. Since we only investigate this source up until the time it makes the transition from the intermediate state to the soft state, our data only show one half of this cycle.

\subsection{RMS spectra}

\subsubsection{Overview}

RMS, or Fourier-resolved, spectra allow us to distinguish varying components of a spectrum (within the frequency range being investigated) from constant components, or those varying at a different frequency. Here we derive rms spectra to attempt to determine the relevance of reflection components within our spectra. As in the previous section, the present data are of sufficient quality to allow us to investigate two frequency ranges, 5-500 mHz (low) and 500 mHz - 5 Hz (high).

\subsubsection{The soft rms spectra of Cygnus X-1}

\begin{figure*}
	\centering
        \includegraphics[width=150mm]{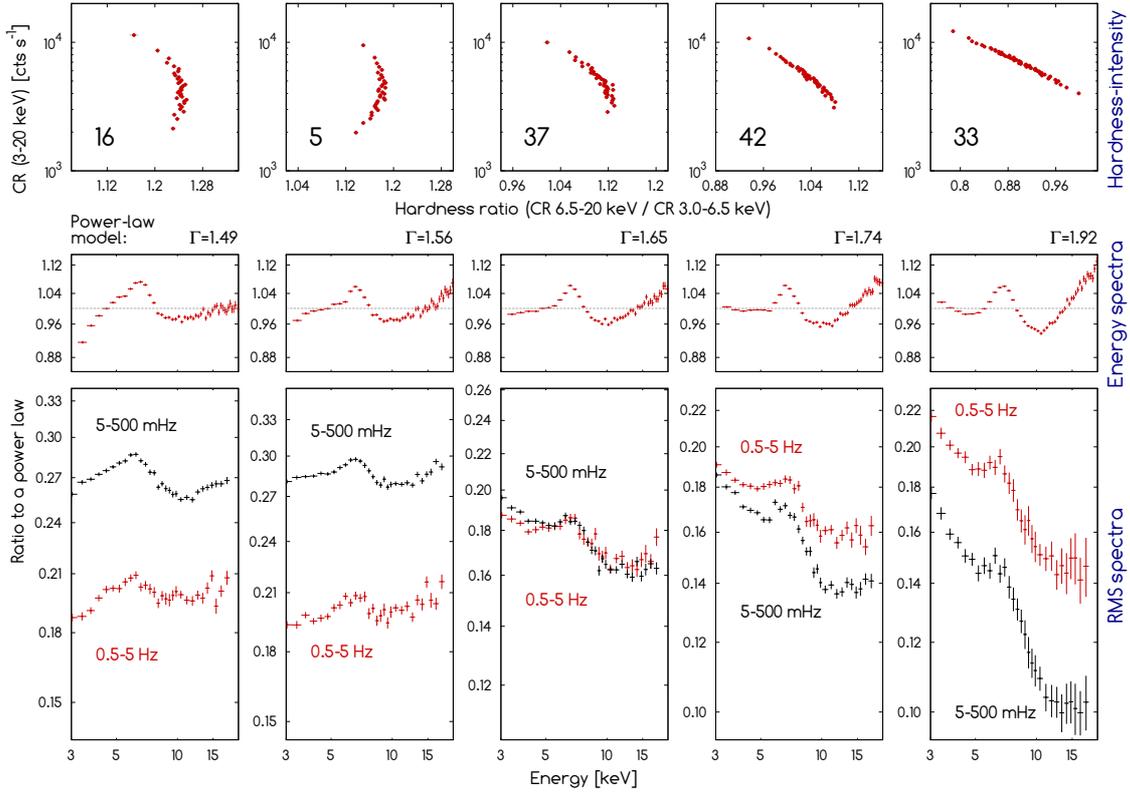}
        \caption{Short time-scale hardness-intensity diagrams (top row), observation-averaged energy spectra (second row) and rms spectra (bottom row) for five observations of Cygnus X-1, sorted (left to right) by decreasing HR, and going from the hard state (obs. 16 and 5) to the hard-intermediate state (obs. 37, 42 and 33). The energy spectra of each observation have been fitted with a single power-law component (the photon index of which is indicated in the panel), and both the energy and rms spectra are shown as ratios to the same power-law model. In the (harder) low-$\Gamma$ observations the hardness-intensity diagrams are of a crescent shape (obs. 5 and 16), and the energy spectra are consistent with a single power-law plus iron line fit; the spectral slopes of the rms spectra from these observations appear similar to those of the energy spectra. As the spectra soften the hardness-intensity diagrams show only strong softer-when-brighter behaviour (obs. 42 and 33), and the energy spectra indicate the presence of separate continuum components in the 3-5 and 10-20 keV bands. For these observations the rms spectra are softer than the energy spectra, and the hard spectral component that dominates the latter above $\simnot10$ keV appears to be strongly suppressed in the former. We also note that as the source softens the fractional rms of the low frequency (5-500 mHz) variability decreases in strength until finally it is dominated in the softest observations by the high frequency (0.5-5 Hz) variability.}
        \label{fig-cygx-1}
\end{figure*}

Fig. \ref{fig-cygx-1} shows the HIDs, the observation-averaged energy spectra and the rms spectra of five observations of Cygnus X-1, which are sorted by decreasing HR. For each observation the energy spectrum was fitted with a single power-law, and the energy and rms spectra are shown as ratios to the same power-law model.

In the observations with the hardest overall energy spectra, the variability in the low frequency band is greater than in the high frequency band, and both bands are dominated by a relatively hard power-law. However, in the observations with the softer energy spectra high frequency variations dominate, and both frequency bands exhibit a steep power-law spectrum.

\subsubsection{The hard component of Cygnus X-1: a result of reflection?}

The rms of any reflected component is expected to be suppressed in comparison to the underlying power-law due to smearing from the light-crossing time of the reflector \citep{Gilfanov2000}, and the hard spectral component could therefore be the result of strong reflected emission. The effect of smearing should be stronger on short time-scales than long time-scales, and we would therefore expect to see greater suppression of the reflected power-law component in the high-frequency rms spectra than in the low-frequency rms spectra.

We have tested our rms spectra for reflection by fitting the PL+R and PL+PL models to the high- and low-frequency rms spectra of obs. 33, which is both the softest and brightest amongst our observations of Cygnus X-1, and its energy spectra show the clearest indication of a change in spectral slope at $\simnot$10 keV. A summary of the fit parameters, and flux contributions from each component, is provided in Table \ref{tbl-rms-fit-parameters}.

Whilst fitting the rms spectra with the PL+R model we chose to fix the power-law photon index ($\Gamma$), the disc normalisation ($K$), the temperature at the inner disc radius ($T_{in}$) and the iron line width ($\sigma$) and centroid energy to the values obtained from fitting the observation-averaged energy spectra, but allowed the \textsc{pexrav} reflection scaling factor to vary. The disc component was convolved with \textsc{cflux} in order that the flux could vary whilst the temperature and inner radius remained fixed. For the PL+PL model we similarly fixed the photon indices of the two power-laws ($\Gamma_{1}$ and $\Gamma_{2}$ respectively) and the iron line width and centroid energy.

\setlength{\tabcolsep}{0.3em}
\begin{table}
	\caption{A summary of the fit parameters for the energy and rms spectra of obs. 33, fitted with the PL+PL and PL+R models. The flux contribution of each component is shown in brackets as a percentage of the overall flux.}
	\label{tbl-rms-fit-parameters}
	\begin{center}
	\begin{tabular}{p{1.6cm}|rl|rl|rl}
		\hline
		& \multicolumn{2}{c}{Energy} & \multicolumn{2}{c}{5-500 mHz rms} & \multicolumn{2}{c}{0.5-5 Hz rms} \\
		& \multicolumn{2}{c}{spectra} & \multicolumn{2}{c}{spectra} & \multicolumn{2}{c}{spectra} \\
		\hline
		\multicolumn{7}{l}{{\bf PL+PL model}} \\[1.5ex]
		\multicolumn{7}{l}{Fit parameters:} \\
		$\chi^{2}_{\rm red}$\dotfill & \multicolumn{2}{c}{1.76 (40 d.o.f.)} & \multicolumn{2}{c}{0.58 (27 d.o.f.)} & \multicolumn{2}{c}{0.33 (27 d.o.f.)} \\
		$\Gamma_{1}$\dotfill & \multicolumn{2}{c}{2.64} & \multicolumn{2}{c}{2.64 $^{f}$} & \multicolumn{2}{c}{2.64 $^{f}$} \\
		$\Gamma_{2}$\dotfill & \multicolumn{2}{c}{1.30} & \multicolumn{2}{c}{1.30 $^{f}$} & \multicolumn{2}{c}{1.30 $^{f}$} \\[1.5ex]
		\multicolumn{7}{l}{Flux contribution [$\times 10^{-11}$ erg cm$^{-2}$ s$^{-1}$]:} \\
		P-L 1 $^{b}$\dotfill & 849.11 & (53.0\%) & 149.19 & (74.1\%) & 176.14 & (65.1\%) \\
		P-L 2 $^{b}$\dotfill & 707.07 & (44.1\%) & 44.90 & (22.3\%) & 85.90 & (31.7\%) \\
		Iron line\dotfill & 46.12 & (2.9\%) & 7.14 & (3.6\%) & 8.71 & (3.2\%) \\[2.2ex]
		\multicolumn{7}{l}{{\bf PL+R model}} \\[1.5ex]
		\multicolumn{7}{l}{Fit parameters:} \\
		$\chi^{2}_{\rm red}$\dotfill & \multicolumn{2}{c}{1.24 (44 d.o.f.)} & \multicolumn{2}{c}{12.51 (26 d.o.f.)} & \multicolumn{2}{c}{2.72 (26 d.o.f.)} \\
		$\Gamma$\dotfill & \multicolumn{2}{c}{2.07} & \multicolumn{2}{c}{2.07 $^{f}$} & \multicolumn{2}{c}{2.07 $^{f}$} \\
		$kT_{\rm in}$ [keV]\dotfill & \multicolumn{2}{c}{0.358} & \multicolumn{2}{c}{0.358 $^{f}$} & \multicolumn{2}{c}{0.358 $^{f}$} \\
		$ref_{\rm refl}\,^{a}$\dotfill & \multicolumn{2}{c}{1.13} & \multicolumn{2}{c}{0.00} & \multicolumn{2}{c}{0.00} \\[1.5ex]
		\multicolumn{7}{l}{Flux contribution [$\times 10^{-11}$ erg cm$^{-2}$ s$^{-1}$]:} \\
		P-L $^{b}$\dotfill & $1\,312.50$ & $(81.1\%)$ & $210.60$ & $(95.8\%)$ & $269.47$ & $(96.9\%)$ \\
		Reflection\dotfill & $272.14$ & $(16.8\%)$ & $0.00$ & $(0.0\%)$ & $0.00$ & $(0.0\%)$ \\
		Disc\dotfill & $8.88$ & $(0.5\%)$ & $4.67$ & $(2.1\%)$ & $3.97$ & $(1.4\%)$ \\
		Iron line\dotfill & $25.99$ & $(1.6\%)$ & $4.53$ & $(2.1\%)$ & $4.58$ & $(1.7\%)$ \\
		\hline
	\end{tabular} \\
	\end{center}
	\textsc{Notes}: $^{a}$ \textsc{pexrav} reflection scaling factor, $^{b}$ Power-law component, $^{f}$ fixed parameter \\
\end{table}

Whereas the PL+R model provided a good fit to the energy spectrum ($\chi^{2}_{\rm red}$ = 1.24), the quality of the fit to the high-frequency rms spectrum was poor ($\chi^{2}_{\rm red}$ = 2.72), and the model failed to provide a satisfactory fit to the low-frequency rms spectrum ($\chi^{2}_{\rm red}$ = 12.51). The fits to the low- and high-frequency rms spectra could be improved dramatically if $\Gamma$ were allowed to vary ($\chi^{2}_{\rm red}$ = 0.28 and 0.30 respectively), but the corresponding increase in $\Gamma$ (from 2.07 to 2.43 and 2.32 respectively) is contrary to our expectation that $\Gamma$ should be equal in the energy and rms spectra. When, conversely, we attempted to fit the observation-averaged energy spectra of obs. 33 with a steeper power-law ($\Gamma = 2.3$) we found that the PL+R model was unable to successfully fit the hard spectral component that dominates above $\rm\simnot10~keV$. Although we note from our spectral fits of Figs. \ref{fig-cyg-soft-excess} and \ref{fig-cyg-17-33-components} that there are some residuals present in the vicinity of the iron line, and these residuals could potentially affect our fits to the energy spectra, we do not believe that they are sufficient to explain our inability to fit the PL+R model to the rms spectra.

Furthermore, attributing the absence of the hard component from the low-frequency (5-500 mHz) rms spectrum to suppression of variability in the reflected emission would require a reflector of at least two light-seconds in diameter (roughly $2.7 \times 10^{4}~R_{\rm G}$), which is much larger than the expected value of $\simnot150~R_{\rm G}$ \citep{Revnivtsev1999} for Cygnus X-1 in the hard state.

The results from fitting the rms spectra with the PL+PL model were more encouraging, with $\chi^{2}_{\rm red}$ less than 1 for both the low- and high-frequency spectra. The strength of the hard power-law relative to the soft power-law is weakened in comparison to the energy spectrum fit, implying that the hard power-law is less variable than the soft power-law, with the suppression stronger at low frequencies than at high frequencies. Although the fits to the energy spectra were slightly poorer than those using the PL+R model ($\chi^{2}_{\rm red} = 1.76$ and 1.24 respectively) there is some indication (from fitting obs. 33; see Section \ref{sec-fitting-additional-component}) that the inclusion of a reflection component (or components) could considerably improve the fit.

\subsubsection{Fractional rms in the high- and low-frequency bands}

We note that in the hard-state observations of Cygnus X-1 (Fig. \ref{fig-cygx-1}, obs. 5 and 16) the fractional rms in the low-frequency band ($f_{\rm rms} \approx$ 25-29 per cent) is greater than that of the higher frequency band ($f_{\rm rms} \approx$ 18-21 per cent). However, in the hard-intermediate-state observations (37, 42 and 33) this situation becomes reversed, and the fractional rms of the high-frequency band is now greater than or equal to that of the low frequency band. This tendency for the high-frequency variability to dominate the rms spectra of the softest observations is consistent with the findings of \cite{Grinberg2014}, who showed that higher frequencies tend to dominate the power spectra as the photon index of the soft power-law increases.

\subsubsection{GX 339-4}

\begin{figure*}
	\centering
        \includegraphics[width=150mm]{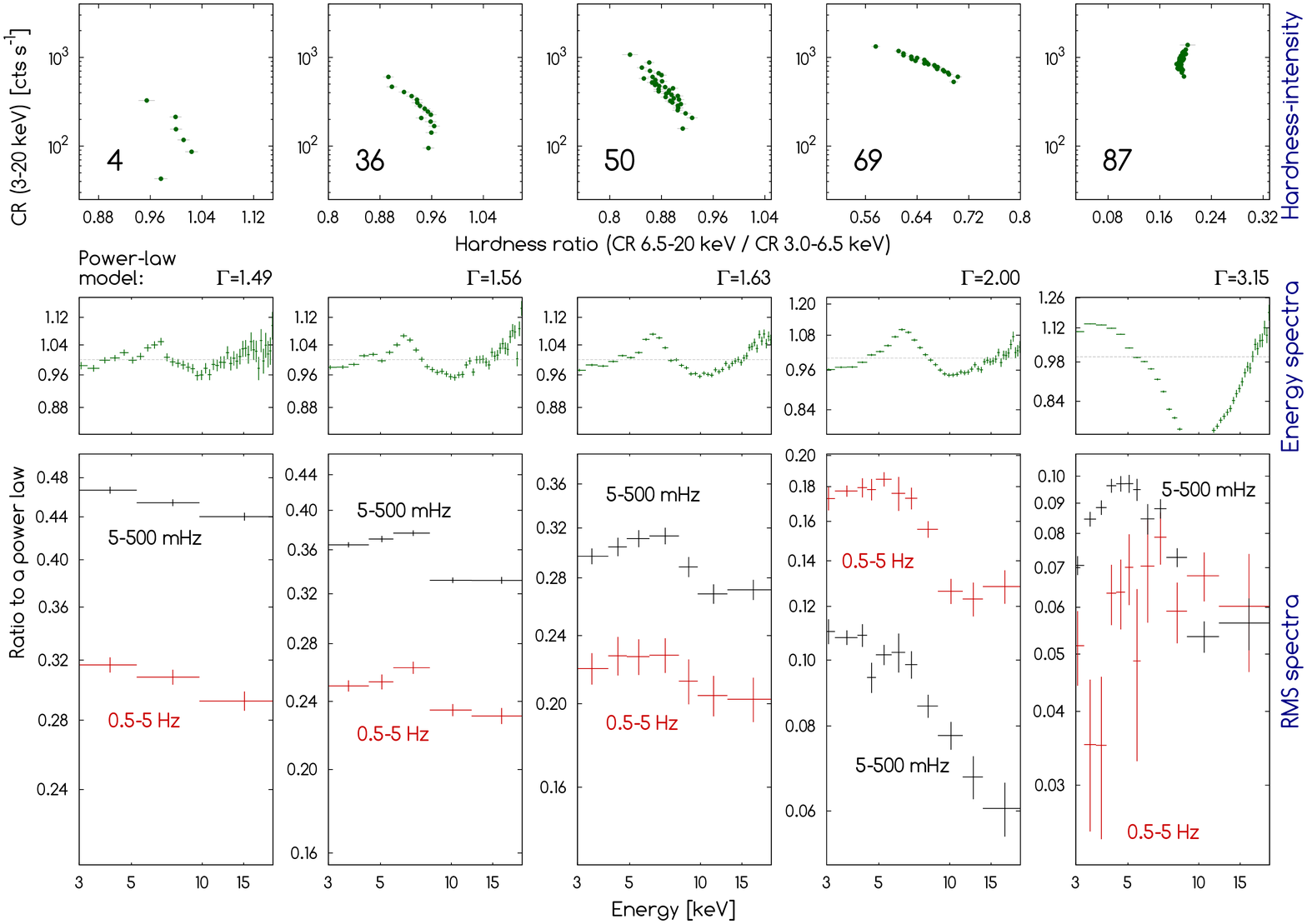}
        \caption{Hardness-intensity diagrams (top row), energy spectra (second row) and rms spectra (bottom row) for five observations of GX 339-4, sorted by decreasing HR. As in Fig. \ref{fig-cygx-1}, the energy and rms spectra of each observation are plotted as a ratio to a power-law. Obs. 4, 36 and 50 are hard state, obs. 69 is hard-intermediate state and obs. 87 is soft-intermediate state.}
        \label{fig-gx339}
\end{figure*}

The evolution of the rms spectra of GX 339-4 (Fig. \ref{fig-gx339}) through the hard and hard-intermediate states (obs. 4 through 69) is very similar to that of Cygnus X-1 (Fig. \ref{fig-cygx-1}), although a possible increase in EW of the iron line with increasing count rate is more pronounced in this source.

However, in obs. 87 we find that GX 339-4 has entered the soft-intermediate state, which we do not see in our Cygnus X-1 data, and in this state a strong thermal disc component is present. In this observation the rms spectrum in the high-frequency band is harder than the energy spectrum due to the suppression of the disc component in the former. The lack of variability in the HID demonstrates the behaviour we would expect to see when an accretion disc dominates the spectrum below $\rm\simnot10~keV$, and is in clear contrast to the large variability in hardness ratio found in the hard-intermediate state observations of Cygnus X-1 (Figs. \ref{fig-hr-cyg-short-term} and \ref{fig-cygx-1}).

\subsubsection{The soft rms spectra of the hard-intermediate state}
\label{sec-soft-rms-spectra}

\begin{figure}
	\centering
        \includegraphics[width=82mm]{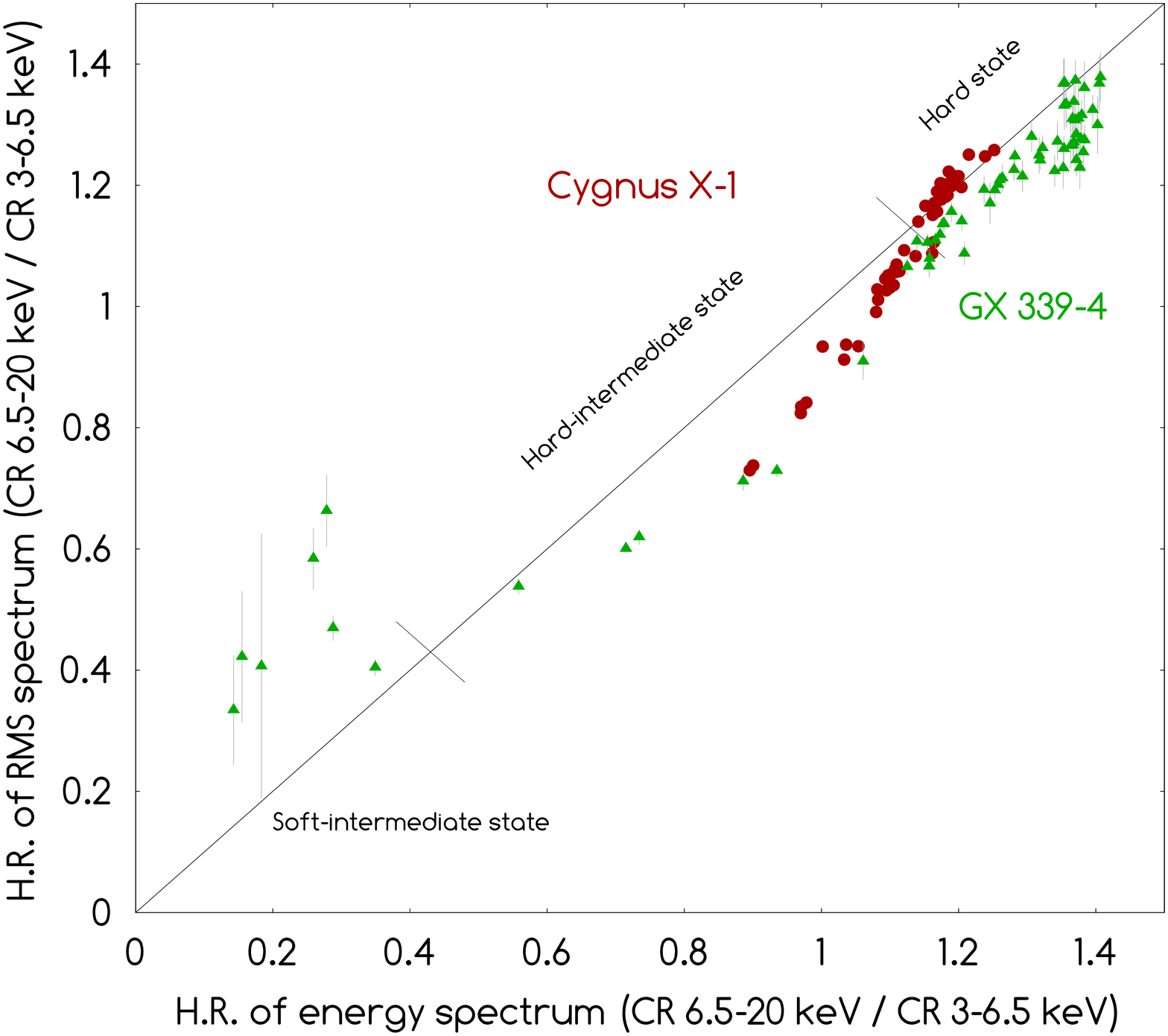}
        \caption{HR of the energy spectra plotted against the HR of the rms spectra (in the 0.5-5 Hz band) for Cygnus X-1 (red circles) and GX 339-4 (green triangles). XTE J1118+480 is too faint to reliably measure the HRs of the rms spectra and has therefore been excluded. In the hard state the energy and rms spectra of Cygnus X-1 and GX 339-4 have similar HRs, but after the transition to the hard-intermediate state the rms spectra become softer than the energy spectra. As GX 339-4 approaches the soft state the spectra start to become dominated by emission from the disc, and the rms spectra now appear harder then the energy spectra. The hard, hard-intermediate and soft-intermediate states are labelled using the hardness intensity diagram of GX 339-4 (Fig. \ref{fig-hr-long-term}, right) as a guide.}
        \label{fig-hr-spec-rms}
\end{figure}

One common feature of the Cygnus X-1 and GX 339-4 spectra shown in Figs. \ref{fig-cygx-1} and \ref{fig-gx339} is the tendency for the rms spectra to become softer than the energy spectra as the count rate increases and the source moves into the hard-intermediate state. This trend only reverses in GX 339-4 once the source moves towards the soft state and a strong thermal disc component enters the spectrum. Fig. \ref{fig-hr-spec-rms} illustrates this point by plotting the HR of the energy spectra (${HR}_{\rm energy}$) against the HR of the rms spectra (${HR}_{\rm rms}$). In the hard state we find that ${HR}_{\rm energy} \approx {HR}_{\rm rms}$, but as the spectra soften and enter the hard-intermediate state we find that ${HR}_{\rm energy} > {HR}_{\rm rms}$; this difference is greatest when ${HR}_{\rm energy} \approx 0.9$, at which point ${HR}_{\rm rms} \approx 0.7$.

\subsubsection{The hard rms spectra of the soft-intermediate state}

In contrast, the soft-intermediate-state observations of GX 339-4 (Fig. \ref{fig-gx339}, obs. 87) reveal the growing influence of the accretion disc on the energy spectra, which are characterised by thermal disc emission in the soft band and a power-law in the hard band. The short-time-scale HIDs show little variability in either HR or CR, and the fractional rms of these data is now much lower than that found in the hard-intermediate state ($< 10$ per cent, with the reduction strongest in the high frequency band). As the disc is more stable than the power-law in the soft state \citep{Churazov2001}, the power-law is relatively stronger (compared to the disc) in the rms spectra than the energy spectra, and we find that ${HR}_{\rm energy} < {HR}_{\rm rms}$.

\section{Discussion}

\subsection{Summary of results}

In order to understand the reason for the hard-state spectral transition at $\dot{m}_{\rm crit}$ we choose to focus our analysis and discussion mainly upon Cygnus X-1, which has an accretion rate which is usually very close to this critical accretion rate. The most relevant results can be summarised as follows:

\begin{itemize}
	\item In the \textbf{hard state} of Cygnus X-1 (obs. 1-24) both the observation-averaged energy spectra (Fig. \ref{fig-hr-cyg-short-term}, top) and rms spectra (Fig. \ref{fig-cygx-1}) consist of a single power-law and iron line, and the short time-scale HIDs (Fig. \ref{fig-hr-cyg-short-term}, bottom) show a clear transition from harder-when-brighter to softer-when-brighter as $\dot{m}$ increases above $\dot{m}_{\rm crit}$.
	\item In the \textbf{hard-intermediate state} of Cygnus X-1 (obs. 25-52) many of the observation-averaged energy spectra (Fig. \ref{fig-hr-cyg-short-term}, top) exhibit a different spectral slope below the iron line ($\rm\simnot3-5~keV$) to that found above the iron line ($\rm\simnot10-20~keV$), and require an additional harder model component such as an additional power-law (the PL+PL model) or a reflection component (the PL+R model). However, the rms spectra (Fig. \ref{fig-cygx-1}) show little or no evidence of this additional component, and appear to be single power-law distributions. The rms spectra are softer than the energy spectra, with photon indices that appear closer to those of the soft component of the energy spectra than to those of the hard component. The short time-scale HIDs (Fig. \ref{fig-hr-cyg-short-term}, bottom) are strongly softer-when-brighter, and no $\dot{m}_{\rm crit}$ spectral transition is observed when the count rate goes below the level at which this transition occurs in the hard state.
	\item Although spectral fitting of the energy spectra of Cygnus X-1 failed to determine which of our two models (PL+PL and PL+R) is the most appropriate choice for these data (see Section \ref{sec-spectral-fitting}), our inability to fit the PL+R model to the low-frequency rms spectrum of obs. 33 without a large increase in $\Gamma$ favours the PL+PL model.
	\item The rms spectra of Cygnus X-1 (Fig. \ref{fig-cygx-1}) are dominated by low-frequency (500 mHz to 0.5 Hz) variability in the hard state, but the fractional rms in the low-frequency band weakens as the source softens, until eventually the high-frequency (0.5 to 5 Hz) variability dominates in the hard-intermediate state.
	\item GX 339-4, for which the accretion rate is greater in all of our observations than that at which the $\dot{m}_{\rm crit}$ transition is observed in Cygnus X-1, shows only softer-when-brighter behaviour (Fig. \ref{fig-hr-gx339-short-term}). Similarly, XTE J1118+480 (for which the accretion rate is less than that of the Cygnus X-1 $\dot{m}_{\rm crit}$ transition) shows only harder-when-brighter behaviour (Fig. \ref{fig-hr-gx339-short-term}). For these sources there appears very little difference in the way that HR (or $\Gamma$) responds to changes in count rate on long and short time-scales (Fig. \ref{fig-edd-ratio-v-gamma}), which is contrary to what we observed in Cygnus X-1.
\end{itemize}

\subsection{Cygnus X-1 across the $\dot{m}_{\rm crit}$ boundary}

The hard-state spectral transition from harder-when-brighter to softer-when-brighter behaviour at $\dot{m}_{\rm crit}$ is observed in Cygnus X-1 (Figs. \ref{fig-hr-cyg-short-term} and \ref{fig-edd-ratio-v-gamma}) on time-scales that are up to the typical observation length of 1-2 ks; this transition is typically observed at count rates of around $\rm5,000~cts~s^{-1}$. In the hard state the rms spectra (Fig. \ref{fig-cygx-1}) show no evidence of greater variability in either the soft or hard bands, which is contrary to what we might expect if a fixed soft component, such as a cool disc or thermal emission, were responsible for causing the harder-when-brighter behaviour at low count rates.

In the hard-intermediate state we do not observe any clear transition to harder-when-brighter behaviour, even though the range of count rates found in some of these observations (i.e. obs. 28 and 51) extend below the point at which a transition to harder-when-brighter behaviour was observed in the hard state. Some hard-intermediate state observations (i.e. obs 37, 39 and 50) do show a slight indication of hardening at lower count rates (below about $\rm3,000~cts~s^{-1}$), and if the emission process were less efficient in the hard-intermediate state we might argue that these lower count rates implied a higher absolute accretion rate than for the same count rates in the hard state. However, since we expect the hard-intermediate state to have softer spectra and a greater radiative efficiency than the hard state then the potentially lower count rates at which the $\dot{m}_{\rm crit}$ transition is observed in the former cannot possibly correspond to the same absolute accretion rate in the latter. The rms spectra (Fig. \ref{fig-cygx-1}) show that in the hard-intermediate state the variability of the source is dominated by the soft spectral component, and the hard component changes comparatively little on either long or short time-scales.

\begin{figure}
	\centering
        \includegraphics[width=82mm]{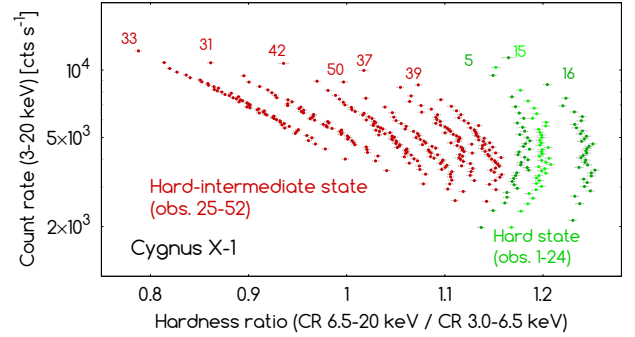}
        \caption{Short time-scale hardness-intensity diagrams for nine observations of Cygnus X-1, separated into hard-intermediate state (obs. 25-52; red points) and hard-state (obs. 1-24; green points). Harder-when-brighter behaviour is only found when the hardness ratio is greater than $\simnot 1.15$, and the count rate is less than $\simnot 5 \times 10^{3}$.}
        \label{fig-hr-multiple-obs}
\end{figure}

We note from Figs. \ref{fig-hr-cyg-short-term} and \ref{fig-edd-ratio-v-gamma} that at low count rates all of the HIDs (in both the hard and hard-intermediate states) appear to approach a single point, which is located at approximately $\Gamma = 1.6, L_{\rm X/E} = 10^{-3}$ (where the luminosity was measured in the 3-20 keV range). This intersection of the HIDs occurs at a luminosity at which the soft spectral component is weak or absent, and only the hard component remains.

We also note from Fig. \ref{fig-edd-ratio-v-gamma} (left) that it may be possible to describe the harder-when-brighter behaviour of both Cygnus X-1 and XTE J1118+480 with the same linear fit, implying a similarly simple, single power-law origin for the harder-when-brighter behaviour of both sources. However, the dynamic range of the Cygnus X-1 data below $\dot{m}_{\rm crit}$ is too low for such a claim to be firmly established.

\subsection{Unifying the hard and hard-intermediate states of Cygnus X-1}

In this paper we have examined the spectra of the hard-intermediate-state Cygnus X-1 observations, and found that a two-power-law interpretation (the PL+PL model) fits the data better. This interpretation agrees with that of \cite{Makishima2008} and \cite{Yamada2013}. A similar model, consisting of a weakly absorbed and highly variable soft component, and a more highly absorbed, slowly varying hard component, was successfully applied by \cite{Noda2014} to describe their \textit{Suzaku} observations of the Seyfert galaxy NGC 3227, and we note the similarity between the variability properties of these components and those described in this paper.

In the hard state, the short time-scale HIDs (Fig. \ref{fig-hr-cyg-short-term}) show a change from harder-when-brighter to softer-when-brighter behaviour as the count rate increases. A simple explanation is that there are two components in the emission, such as the Comptonisation of thermal seed photons from the accretion disc and the Comptonisation of cyclo-synchrotron seed photons from the RIAF and corona itself, and we are observing a change in the relative proportion of these two components. A similar model was proposed by \cite{Skipper2013} to explain the complex correlations between spectral index and count rate on very short time-scales.

In the hardest of the hard-intermediate state observations (e.g. obs. 28 and 35, Fig. \ref{fig-hr-cyg-short-term}, and obs. 50, 37 and 39, Fig. \ref{fig-hr-multiple-obs}) there is a hint of harder-when-brighter behaviour, possibly at a slightly lower count rate than in the hard state, but most of the hard-intermediate-state observations are at higher count rates and show only softer-when-brighter behaviour, consistent with disc-dominated seed photon supply. One possible explanation for the lower count rates of the $\dot{m}_{\rm crit}$ transition in the hard-intermediate state is that the accretion disc persists to smaller radii, and thus the harder-when-brighter behaviour which is characteristic of domination by seed photons from the corona or RIAF does not dominate until lower absolute accretion rates.

The tendency for both Cygnus X-1 and GX 339-4 to become increasingly dominated by high-frequency variability as the source softens through the hard and hard-intermediate states (Fig. \ref{fig-cygx-1}) is consistent with the inward motion of a truncated accretion disc, although the low-energy limit of 3 keV precludes us from attempting to fit any cool disc component which may be present in the hard state. The observations are furthermore consistent with variations being generated mainly in the disc, with higher frequencies coming from smaller radii \cite[e.g.][]{Lyubarskii1997, Arevalo2006}.

Whereas the rms spectra are reasonably well fitted by a single power-law in both the hard and hard-intermediate states, the energy spectra in these states are slightly different. In the hard state the energy spectra are fitted well by a single power-law with slope similar to that of the rms spectra, but the short time-scale HIDs (Fig. \ref{fig-hr-cyg-short-term}) show some harder-when-brighter behaviour and some softer-when-brighter behaviour, indicative of two components. The single power-law fits to both the energy and rms spectra may simply mean that both components have similar spectral indices, arising from similar optical depths and temperature distributions. However, one component could still be more variable than the other.

In the hard-intermediate state an additional power-law is required to account for the excess emission in the energy spectrum above 10 keV. The spectral fits show that a second power-law, rather than a reflection component, is the best fit to this component, which we therefore associate with Comptonisation of seed photons from the RIAF. This hard component is considerably weakened in the low-frequency rms spectra (Fig. \ref{fig-cygx-1}), and is absent from the high-frequency rms spectra. Overall fractional variability is lower in the hard-intermediate state than in the hard state (Fig. \ref{fig-cr-rms}), and therefore this component must be less variable than the soft power-law component arising from disc variations.

The dual power-law spectra found in the hard-intermediate state suggest a much-increased difference between the temperature (due to the increasing effect of cooling from the inwardly-moving disc) and/or the optical depth \cite[e.g.][]{Makishima2008} of the inner accretion flow (where the cyclo-synchrotron seed photons are Comptonised) and the outer accretion flow (where the disc seed photons are Comptonised).

We note that a strongly polarised hard tail, possibly from a jet, has been detected above 400 keV in Cygnus X-1 using the \textit{INTEGRAL/IBIS} telescope \citep{Laurent2011}. However, the polarised hard tail is not bright enough to explain the hard component seen here at lower energies. Extrapolating the polarised hard tail down to lower energies indicates that it would only contribute $\simnot6 - 12$ per cent of the $\rm3 - 20~keV$ flux seen here; alternatively, extrapolating the hard component deduced from obs. 33 (Table \ref{tbl-rms-fit-parameters}) to $\rm400 - 1000~keV$ indicates that it would be $\simnot25$ times brighter than the observed polarised hard tail. We also note that the polarised hard tail has so far been seen only in a significantly softer state than those which we have discussed here, and so, at this stage, we do not associate it with the hard component seen here at lower energies.

\section{Conclusion}

We conclude that most of the variability properties discussed in this paper can be explained by the combination of a soft, variable component driven by disc variations with seed photons coming from the disc, and a second, harder, but less variable component arising from cyclo-synchrotron seed photons from the RIAF or corona. We associate the single harder-when-brighter power-law of XTE J1118+480 with the hard power-law found in Cygnus X-1, and the softer-when-brighter power-law which dominates the spectra of GX 339-4 with the soft power-law found in Cygnus X-1.

\section{Acknowledgements}

This research has made use of data obtained through the High Energy Astrophysics Science Archive Research Center Online Service, provided by the NASA/Goddard Space Flight Center. CJS acknowledges support from a STFC studentship and a summer bursary from the University of Southampton. IM\textsuperscript{c}H acknowledges support from STFC under grant ST/G003084/1.

\bibliographystyle{kp}
\bibliography{references}

\label{lastpage}
\end{document}